\documentclass{article}
\usepackage{amsmath}
\usepackage{amsfonts}
\usepackage{amssymb}
\usepackage{graphicx}
\usepackage{authblk}
\usepackage{tikz}

\usepackage{program}

\newtheorem{theorem}{Theorem}
\newtheorem{remark}{Remark}%
\newtheorem{lemma}{Lemma}

\newtheorem{corollary}[theorem]{Corollary}

\def\RR{\mathbb R}

\def\cL{\mathcal{L}}

\def\cQ{\mathcal{Q}}

\def\br{\color{red}\mathbf}
\def\bb{\color{blue}\mathbf}

\title{Deterministic $n$-person shortest path and terminal games
on symmetric digraphs have Nash equilibria in pure stationary strategies}

\begin{document}

\author[1]{Endre Boros}

\author[2]{Paolo Giulio Franciosa}

\author[3]{Vladimir Gurvich}

\author[4]{Michael Vyalyi}

\affil[1]{\texttt{endre.boros@rutgers.edu} MSIS \& RUTCOR, Business School, Rutgers University, 100 Rockafellar Road, Piscataway, 08854, NJ, USA}

\affil[2]{\texttt{paolo.franciosa@uniroma1.it} Department of Statistics, Sapienza University, P.le Aldo Moro 5, Rome, 00185, Italy}

\affil[3]{\texttt{vgurvich@hse.ru} MSIS \& RUTCOR, Business School, Rutgers University, 100 Rockafellar Road, Piscataway, 08854, NJ, USA}

\affil[4]{\texttt{vyalyi@gmail.com} Higher School of Economics (HSE) of National Research University, and  Federal Research Center ``Computer Science and Control'' of the Russian Academy of Sciences, and Department of Control and Applied Mathematics of Moscow Institute of Physics and Technology, Moscow, Russia}

\date{}

\maketitle


\abstract{We prove that a deterministic $n$-person shortest path game has a Nash equlibrium
in pure and stationary strategies 
if it is edge-symmetric 
(that is  $(u, v)$ is a move
whenever  $(v, u)$  is, apart from moves entering terminal vertices) and
the length of every move is positive for each player.
Both conditions are essential, though it remains an open problem
whether there exists a NE-free $2$-person non-edge-symmetric  game with positive lengths.
We provide examples for NE-free $2$-person edge-symmetric games that are not positive.
We also consider the special case of terminal games 
(shortest path games in which only terminal moves have nonzero length, possibly negative)
and prove that edge-symmetric $n$-person terminal games 
always have Nash equilibria in pure and stationary strategies. 
Furthermore, we prove that an edge-symmetric $2$-person terminal game 
has a uniform (subgame perfect) Nash equilibrium, provided 
any infinite play is worse than any of the terminals for both players.
}

\maketitle

\section{Introduction}


Given a finite directed graph $G = (V,E)$,
we interpret a vertex  $v \in V$  as a position of a game and
a directed edge $e = (u,v)\in E$  as a move from  $u$  to  $v$.

The set of players is denoted by  $I = \{1, \ldots, n\}$.
Each player $i\in I$ controls a subset $V_i\subseteq V$ of the positions, and we also have a nonempty subset $V_T$, the set of so-called \emph{terminal} positions that are not controlled by any of the players. 
We assume that the sets $V_1$,..., $V_n$, and $V_T$ form a partition of $V$, and that terminals are the only positions with no directed edges leaving. Furthermore, we fix a position  $v_0\in V\setminus V_T$  and call it the {\em initial} position.

A pure and stationary strategy  $\sigma_i$  of player  $i\in I$  is a mapping
$\sigma_i:V_i\mapsto V$ such that $(v,\sigma_i(v))\in E$ for all $v \in V_i$.
To simplify our notation, sometimes we also use $\sigma_i$ 
as the set of edges $\{(v,\sigma_i(v))\mid v\in V_i\}$. 
In this paper we consider only pure and stationary strategies, 
and call them simply strategies. 

Let us denote by $\Sigma_i$ the set of strategies of player $i\in I$, and set
$\Sigma=\Sigma_1\times\cdots \times \Sigma_n$.
The tuple $\sigma = \left(\sigma_i \mid i \in I\right)\in \Sigma$  is called
a {\em strategy profile} or {\em situation}.
Every situation determines uniquely
a  directed walk, called the \emph{play} $P(\sigma)$ and defined  as follows. 
In a position  $v \in V_i$ on this walk, player  $i$  makes the move  $(v, \sigma_i(v))$, 
after which the walk enters $v'=\sigma_i(v)$. 
Starting with $v_0$, this process creates a unique   
play  $P(\sigma)$.
This play either comes to a terminal $v\in V_T$, where it stops 
(no player handles the terminal and no moves are from it), 
or it repeats a position reached earlier, 
in which case it follows the enclosed cycle infinitely many times.
In the first case we call $P(\sigma)$ a \emph{terminal play}, 
while in the second case it is called an \emph{infinite play}.


Each player $i\in I$ has his own length function 
$\ell^i:E\rightarrow \RR$ that assigns to every move $e\in E$ 
a real number $\ell^i(e)$ that we 
interpret as the cost of this move for this player;  
we call it the $\ell^i$-cost of the move and  set $\ell=(\ell^i\mid i\in I)$. 
The cost of a sequence of moves is the sum of the costs of these moves. 
In particular, for a situation $\sigma\in\Sigma$ and player $i\in I$ the \emph{effective cost} of $\sigma$ for player $i$ is
\[
\ell^i(\sigma) ~=~ 
\sum_{(u,v)\in P(\sigma)} \ell^i(u,v).
\]
This kind of additive effective cost function, 
called \emph{total}, is considered in \cite{TV87}; 
see more analysis and variations in \cite{BEGM14,BEGMO17,GO14}.
Note that the cost of an infinite play may not be well defined, 
in particular when the cost of moves can be both positive and negative. 
The above cited papers provide several possible solutions for this problem. 
In our paper we focus on the special case, 
when all cost functions are positive;  
in this case the cost of an infinite play is $+\infty$, 
while every terminal play has a finite cost. 

All players aim to minimize their effective costs.

\medskip

We will call the obtained class of games
{\em the shortest path games}, or simply games in the sequel.
Such a game $\Gamma$ is defined  
by the triple  $(G, v_0, \ell)$. 
Note that we always assume a given partition 
$V = V_1 \cup \cdots \cup V_n\cup V_T$ of the vertices of $G$; 
just we do not want to convolute our paper with extra notation.




A situation  $\sigma \in \Sigma$  is called
a {\em Nash equilibrium} (or NE, in short)  if
$\ell^i(\sigma) \leq \ell^i(\sigma')$  for  any player  $i \in I$
and any situation  $\sigma' \in \Sigma$  that
may differ from  $\sigma$  only by the strategy of player  $i$.
In other words, a situation is a NE if
no player  $i \in I$  can reduce her effective cost
by choosing another strategy, provided that all other players keep their old strategies. 
We call $\sigma$ a {\em terminal NE} 
if it is a NE and $P(\sigma)$ is terminal.

\medskip

\subsection{Edge-symmetric and positive shortest path games}

A local cost function $\ell$  is called \emph{positive} if
$\ell^{i}(e) > 0$  for  all players  $i \in I$  and moves $e \in E$.
This condition is equivalent to the seemingly weaker
condition of
$\sum_{e \in C} \ell^i(e) > 0$ for all directed cycles  $C$  of  $G$ and players $i \in I$.
The equivalence is based on the concept of potential transformation
introduced for directed  graphs in \cite{Gal58}.  
It is applied to the length function of each player, separately.
Such an equivalent transformation can in fact be computed efficiently, 
by linear programming. 
Since this transformation changes all path lengths between two 
vertices by the same constant, 
a shortest path game with positive cycle lengths can always be transformed into an equivalent game with positive edge lengths.

A digraph $G=(V,E)$ is called \emph{symmetric} if $(u,v)\in E$ whenever $(v,u)\in E$, except if $u\in V_T$.
Finally, we call a game $\Gamma =(G, v_0, \ell)$  {\em edge-symmetric and positive}
if $G$ is symmetric and $\ell$ is positive. Let us emphasize that the cost functions of the players are not assumed to be symmetric, that is $\ell^i(u,v)\neq\ell^i(v,u)$ may hold for  
some edges $(u,v)\in E$ and players $i\in I$. 

\begin{theorem} \label{t1}
An edge-symmetric and positive $n$-person shortest path
game $\Gamma =(G, v_0, \ell)$ has a NE. 
Furthermore, it has a terminal NE whenever $V_T$ is reachable from $v_0$.
\end{theorem}

Let us add that the above result cannot be extended to the so-called limit mean effective payoff games (see \cite{Gil57} for definitions). 
For mean effective payoff a $2$-person NE-free example was constructed in \cite{Gur88} on a symmetric bipartite graph with non-positive length function, and it is easy to make this example positive by adding a constant to the lengths of all edges.

Both conditions of edge-symmetry and positivity,
are essential in Theorem~\ref{t1}.
In Section \ref{NEfree} we provide examples of
NE-free edge-symmetric (but not positive) $2$-person games.
For $n \geq 3$  such examples were known earlier \cite{BG03,GO14}.
Furthermore, a NE-free positive
but not edge-symmetric $3$-person game was obtained in \cite{GO14}.

Another case, when the assumption of symmetry helps
to give criteria for the existence of a NE is the family of
so-called cyclic games, in which each directed cycle is a separate outcome.
For $2$-person edge-symmetric cyclic games \cite{BGMS11} provides a criterion for the existence of NE for all cost functions. Without the assumption of symmetry no such criterion is known. 

In \cite{FSU92}, a polynomial algorithm solving
game ``geography" was found for symmetric digraphs,
while the problem is known to be PSPACE-complete in general \cite{LS80,Sch78}.

Note also that positivity is important even if we have only one player. 
In the presence of negative arc lengths 
(with some negative cycles) a shortest path may not exist, and 
computing a shortest simple path is NP-hard in this case. 

A shortest path game is called play-once 
if $\vert V_i\vert =1$ for all $i\in I$. It was shown in \cite{BG03} that
every play-once positive game has a NE.

It remains an open problem whether every $2$-person 
non-edge-symmetric positive shortest path game has a NE or not.

\medskip

\subsection{Edge-symmetric terminal games}

A related family of games that in fact forms 
a special case of shortest path games is the family of terminal games.
In many popular positional games
(Go, Chess, Checkers, etc.)
players do not pay for moves; the effective cost for each player
depends only on the terminal position.
We assume, as before, that the set of positions and moves of the game form a directed graph, $G=(V,E)$, players $i\in I$ control distinct subsets $V_i\subseteq V$ of the positions, and $V_T$ is a nonempty subset of the terminal positions, such that $V=V_1\cup\cdots\cup V_n\cup V_T$ is a partition of the set of positions. In these games a situation may end up in a terminal position within $V_T$ or in an infinite play, going around the same cycle infinitely many times. We consider all infinite plays equivalent, and denote the corresponding outcome by $c$. Thus, to evaluate the outcome of these games we have mappings  $\cL^i : V_T \rightarrow \RR$ for $i\in I$ that evaluate the value of terminals for players.

Then, for a situation  $\sigma$
that defines a play  $P(\sigma)$ the effective cost of player $i\in I$ is defined as
\[
\cL^i(\sigma) ~=~ \begin{cases}
	\cL^i(w) & \text{ if } P(\sigma) \text{ is a finite play terminating at } w\in V_T,\\
	0 &  \text{ if } P(\sigma) \text{ is the infinite play $c$.}
\end{cases}
\]
The infinite play $c$
may be worse than some terminals and
better than some others for a player. As in shortest path games, all players minimize their own effective costs. We call such a game $\Gamma=(G,v_0,\cL)$ a \emph{terminal game}.

In the $2$-person zero-sum case, such games are also called 
{\em deterministic graphical}; see \cite{Was90}, where 
the existence of a NE (saddle point) in pure and stationary strategies 
was shown.

In  \cite{BG03}, this result was extended to
the non-zero-sum $2$-person case using a general criterion of \cite{Gur75,Gur88}.
Recently these results were extended further
for the so-called multi-stage  deterministic graphical games;
see \cite{Gur18}  and also  \cite{GK18}.
However, this line of arguments cannot be extended to
the $n$-person case for $n>2$, since the
criterion of  \cite{Gur75,Gur88}  holds only for $n=2$.
Examples for $3$- and $4$-person NE-free terminal games were constructed in \cite{BGMOV18,Gur15}.

Let us also mention that Everett \cite{Everett57} 
already in 1957 considered a closely related class of terminal games 
and, among other results, provided some NE-free examples 
for concurrent zero-sum terminal games. 

The existence of a NE for the above defined class of terminal games remains an open problem
if we assume additionally that $\cL^i(v)< 0$ for all terminals $v\in V_T$ and players $i\in I$.
This condition describes a natural subclass of these games in which an infinite play is worse than any terminal play for each of the players. We can show that this subfamily of terminal games can be viewed as a subfamily of positive shortest path games, and thus the existence of a (terminal) NE for such a game is implied by Theorem \ref{t1}.
We can further generalize this by waving the above assumption, and prove the following claim:

\begin{theorem}\label{t2}
An edge-symmetric $n$-person terminal game has a NE.
\end{theorem}

\medskip

\subsection{Uniform Nash equilibrium for edge-symmetric terminal games}

We can further strengthen Theorem \ref{t2} for the case of $2$ players. We call a situation $(\sigma_i\mid i\in I)$ a \emph{uniform Nash equilibrium} (or a UNE, in short), if it is a NE no matter what the initial position is. 

\begin{theorem}\label{t3}
Assume that $\Gamma=(G,v_0,\cL)$ is a terminal game that satisfies the following conditions:
\begin{description}
	\item[(TWO)] $n=\vert I\vert =2$;
	\item[(SYM)] $G$ is symmetric;
	\item[(CIW)] infinite plays are worse than any of the terminal plays for all players.
\end{description}
Then $\Gamma$ has a UNE.
\end{theorem}

Let us add that all three conditions are necessary to guarantee the existence of a UNE, as we demonstrate it in Section \ref{s-UNE}.

We can also note that condition (CIW) is automatically satisfied by positive shortest path games. However, the theorem still fails even if we keep all three conditions but replace terminal games by shortest path games, see the last example in Section \ref{NEfree}.

\smallskip

\section{Edge-Symmetric Positive Shortest Path Games}

Assume that $\Gamma=(G,v_0,\ell)$ is an edge-symmetric and positive shortest path game. We denote by $N^+(v)=\{u\mid (v,u)\in E\}$ the out-neighborhood of a position $v\in V$ (and thus we have $N^+(v)=\emptyset$ for all terminals $v\in V_T$). Note also that since all plays reaching a terminal end there, we can merge all terminals into a single terminal position without any loss of generality. Thus, in this section we assume that $V_T=\{v_t\}$. 

Let us consider the subgraphs induced by vertex sets $V_i$, $i\in I$. Each such subgraph can uniquely be decomposed into strongly connected subgraphs. We denote by $\cQ=\{Q_1,Q_2,\ldots\}$ the family of all such components for all $i\in I$. For sake of simplicity, we consider $\{v_t\}$ also as one of those components. Note that we adopt the name "component" for these subgraphs, even though they may not be strongly connected components of our graph. They are strongly connected components of the subgraphs, induced by $V_i$, $i\in I$ and $V_T=\{v_t\}$. 

We use $i(v)$ to denote the player who controls vertex $v$. We extend this notation even for the terminal node for notational convenience and assume $i(v_t)\not\in I$, even though $v_t$ has no outgoing arcs, and no player controls it. 
We denote by $i(Q)\in I$ the player who controls the vertices in such a component $Q\in \cQ$. Note that, due to the symmetric nature of the graph and the above definition of the components, we have the following property:
\begin{itemize}
\item[\rm (A)] If $u\in Q$, $v\in Q'$, $Q\neq Q'$ and $(u,v)\in E$ then $i(Q)\neq i(Q')$.
\end{itemize}

Since a component $Q\in\cQ$ is strongly connected, for any two vertices $u,v\in Q$ player $i(Q)$ can create directed path(s) between these vertices. We denote by $d^{i(Q)}(u,v)$ the length of a shortest $u\to v$ path within $Q$, where the $i(Q)$-length is used to measure the length of the edges. Note that we do not assume the symmetry of the lengths, thus $d^{i(Q)}(u,v)$ and $d^{i(Q)}(v,u)$ may be different values.

Let us define $Q(v)\in \cQ$ to be the component containing vertex $v\in V$.

For a $v_0\to v_t$ path $P$ let us call a sequence $\{v_i,v_{i+1},\ldots v_{i+j}\}$ a subpath of $P$. The sets $P\cap Q$, $Q\in\cQ$ partition $P$ into a number of such subpaths. We denote by 
$q(P)$ the number of these subpaths. In other words, if we denote by $Q(P)_j$, $j=1,...,q$ the components that $P$ intersects as we follow $P$ from $v_0$ to $v_t$, then $q=q(P)$, and $Q(P)_j\neq Q(P)_{j+1}$ for all $j=1,...,q-1$. Note that we may have $Q(P)_j=Q(P)_k$ for some $j$ and $k\geq j+2$. 

\subsection{Special paths}

Let us now consider a $v_0\to v_t$ path $P$ in $G$, a player $i\in I$, and the subgraph $G(i,P)$ of $G$ consisting of the moves in $P$ and the moves $(u,v)\in E$ with $u\in V_i$. We say that $P$ is \emph{$i$-special} if $P$ is a shortest $v_0\to v_t$ path in $G(i,P)$ for player $i$.

\begin{lemma}\label{l1}
There exists a $v_0\to v_t$ path $P$ satisfying the following conditions:
\begin{itemize}
\item[\rm (B)] $q(P)$ is the smallest among all $v_0\to v_t$ paths, and
\item[\rm (C)] $P$ is $i$-special for all $i\in I$.
\end{itemize}
\end{lemma}
\begin{proof}
Let us first associate a new edge length $\lambda: E\mapsto \{0,1\}$ by defining
\[
\lambda(u,v) ~=~ \begin{cases}
1 & \text{ if } Q(u)\neq Q(v),\\
0 & \text{ if } Q(u)= Q(v),
\end{cases}
\]
and choose a $\lambda$-shortest $v_0\to v_t$ path $P$. By abusing our notations, we shall use $P$ to denote both the set of vertices of $P$ and also the set of edges in $P$. We hope that from the context it will always be unambiguous what we mean.

Note that $q(P)$ is minimum among all $v_0\to v_t$ paths, and that $P$ does not cross any component $Q\in\cQ$ twice. This is because $Q$ is strongly connected and the $\lambda$-cost of a path inside $Q$ is zero, and outside is positive. Let us observe that the minimality of $q(P)$ implies the following property:
\begin{itemize}
\item[\rm (D)] For all $v\in Q(P)_j$, $1\leq j\leq q(P)-2$ and $j+2\leq k \leq q(P)$ we have $N^+(v)\cap Q(P)_k=\emptyset$.
\end{itemize}
Let us introduce $P[j]=P\cap Q(P)_j$ for $j=1,...,q(P)$ and note that the edges in $P$ leaving vertices in $P[j]$ form a subpath ending in $Q(P)_{j+1}$.
Let us next define
\[
r(P)_j~=~ \sum_{u\in P[j]\atop (u,v)\in P} \ell^{i(Q(P)_j)}(u,v)
\]
for $j=1,...,q(P)-1$, and set $r(P)=(r(P)_1,...,r(P)_{q(P)-1})$. Note that in a way $r(P)$ measures the length of $P$ such that the length of every move is measured by the length function of the player who controls that move.

Assume now that $P$ is an arbitrary path that satisfies property (B). Our arguments above show that there are such paths. Assume also that it does not satisfy property (C), that is there exists a player $i\in I$ that can improve on it. It means that there is an index $1\leq j < q(P)$ with $i=i(Q(P)_j)$ and a position $v\in P[j]$ such that player $i$ can deviate from $P$ at $v$, return to $P$ in $Q(P)_j\cup Q(P)_{j+1}$ (due to property (D)) and make the $i$-length of the improved path $P'$ shorter than the $i$-length of $P$. Note that if $P'$ returns to $P$ at or before the first position of $P[j+1]$, then we must have $r(P')_j<r(P)_j$ and $r(P')_k=r(P')_k$ for all $k>j$. If $P'$ returns to $P$ at or after the second position of $P[j+1]$ (in case $P[j+1]$ has more than one vertex), then we must have  $r(P')_{j+1}<r(P)_{j+1}$ and $r(P')_k=r(P)_k$ for all $k>j+1$. Note that in the second case we may have $r(P')_j>r(P)_j$. In  both cases however, the vector $r(P')$ is smaller than $r(P)$ in the reverse lexicographic order. Note furthermore that $P'$ must also satisfy property (B). 

Since we have only finitely many different paths, and thus $r(P)$ vectors, after finitely many improvement steps we must arrive to a path that satisfies both (B) and (C), as claimed.
\end{proof}

Let us call a path $P$ satisfying both (B) and (C) a \emph{special path}. By the above lemma such a special path exists. Let us now fix one $v_0\to v_t$ special path $P$.

\subsection{Extending a special path to a NE}

Let us introduce for $j=1,...,q(P)$ the sets
\[
U(P)_j ~=~ \bigcup_{k=1}^j Q(P)_k,
\]
and let $u_j\in Q(P)_j$ be the first vertex on $P\cap Q(P)_j$. Recall that for a vertex $u\in Q(P)_j$ and player $i=i(Q(P)_j)$ we denote by $d^i(u_j,u)$ the $i$-distance from $u_j$ to $u$ inside component $Q(P)_j$.

We are ready now to define strategies for the players.
\begin{itemize}
\item[(i)] If $v\in P$ then we choose edge $(v,u)\in P$.
\item[(ii)] If $v\in V\setminus P$ and $N^+(v)\cap U(P)_{q(P)}\neq\emptyset$, then
\begin{itemize}
\item[(ii-1)] choose the smallest index $k$ such that $N^+(v)\cap Q(P)_k\neq\emptyset$ and set $i=i(Q(P)_k)$;
\item[(ii-2)] choose a vertex $u\in N^+(v)\cap Q(P)_k$ that minimizes $d^i(u_k,u)$;
\item[(ii-3)] choose edge $(v,u)$.
\end{itemize}
\item[(iii)] If $v\in V\setminus P$ and $N^+(v)\cap U(P)_{q(P)}=\emptyset$, then choose an arbitrary edge $(v,u)\in E$.
\end{itemize}
Let us denote by $\sigma(P)=\{\sigma_i(P)\mid i\in I\}$ one of the situations defined in this way.

\begin{lemma}\label{l2}
If $P$ is a special $v_0\to v_t$ path, then $\sigma(P)$ is a NE.
\end{lemma}

\begin{proof}
Assume for contradiction that a player $i\in I$ can deviate and improve on $P$. This means that there is a component $Q(P)_j$ controlled by $i=i(Q(P)_j)$ and there is a vertex in $P\cap Q(P)_j$ such that player $i$ can deviate from $P$ starting at this vertex and create a new path $P'$ such that the $i$-length of $P'$ is smaller than the $i$-length of $P$. We can assume that $P'$ is the shortest path (according to $\ell^i$) player $i$ can create, and that $P\cap Q(P)_k=P'\cap Q(P)_k$ for all $k<j$.

In particular, $P'$ cannot return to $P\cap U(P)_{j-1}$. Note first that if $u\in U(P)_{j-1}\setminus P$, $i(u)\neq i$, and $(u,v)\in \sigma(P)$, then we must have $v\in U(P)_{j-1}$ by our definition of $\sigma$, since our graph is symmetric. Note next that if $u\in U(P)_{j-1}$, and $i(u)=i$, then $u\in U(P)_{j-2}$. These together imply that if $P'$ enters $U(P)_{j-1}$ then it cannot reach $v_t$.

Let us now focus on the segment of $P'$ that is outside of $U(P)_{j-1}$ (note that this may be the entire path $P'$ if $j=1$.)
Let us denote the vertices along this segment of $P'$ by $u_j=w_0$, $w_1$, ..., $w_m=v_t$. Let $k$ be the smallest index such that $w_k\not\in Q(P)_j$. By property (C) and the fact that $P'$ is the shortest path that player $i$ can create, we can conclude that $w_k\not\in P$. Then we must have $i(w_k)\neq i$ by the definition of the components, thus we have $(w_k,w_{k+1})\in \sigma(P)$. By the symmetric nature of our graph, by part (ii) of our definition of $\sigma$, and by the fact that $P'$ does not enter $U(P)_{j-1}$ we must have $w_{k+1}\in Q(P)_j$. Since $N^+(w_k)\cap Q(P)_j\supseteq \{w_{k-1},w_{k+1}\}$, we get that (ii-2) of the definition of $\sigma$ implies $d^i(u_j,w_{k+1})\leq d^i(u_j,w_{k-1})$. Since player $i$ can create such a shortest $u_j\to w_{k+1}$ path inside $Q(P)_j$ he could replace $w_0,...,w_{k+1}$ with this path, and create another path $P''$ that is strictly shorter (by at least $\ell^i(w_{k-1},w_k)+\ell^i(w_k,w_{k+1})$) than $P'$, contradicting the fact that $P'$ is the shortest path this player can create deviating from $P$ inside $Q(P)_j$. This contradiction proves our claim.
\end{proof}

\noindent{\textbf{Proof of Theorem \ref{t1}}:}
The claim follows by Lemmas \ref{l1} and \ref{l2}.
\qed

\bigskip

\section{Terminal Games and Shortest Path Games}

It is easy to see that terminal games satisfying condition (CIW) of Theorem \ref{t3} 
can also be viewed as positive shortest path games. 

\begin{theorem}\label{t4-terminal=shortest-path}
	Assume that $\Gamma=(G,v_0,\cL)$ is a terminal game that satisfies condition (CIW). Then there exist positive local costs functions $\ell^i:E\mapsto \RR$ for $i\in I$ such that $\Gamma'=(G, v_0, \ell)$ is a positive shortest path game such that any situation $\sigma$ that is a terminal NE of $\Gamma'$ is also a NE of $\Gamma$. 
\end{theorem}

\begin{proof}
Let us note first that condition (CIW) implies that $\cL^i(v)<0$ for all terminals $v\in V_T$ and players $i\in I$. Since we have only finitely many terminals and players, we can assume w.l.o.g. that $\cL^i(v)$ are all negative integers for all terminals $v\in V_T$ and players $i\in I$, and there exists a positive integer $M$ such that
\[
-M ~<~ \cL^i(v) <0 ~~~\text{ for all } i\in I, \text{ and } v\in V_T.
\]
Let us next define the local costs for the players $i\in I$ and moves $(u,v)\in E$ by
\[
\ell^i(u,v) ~=~ 
\begin{cases}
\frac{1}{2\vert E\vert } & \text{ for all moves } (u,v)\in E, ~v \not\in V_T,\\
M+\cL^i(v) & \text{ for all moves } (u,v)\in E, ~v \in V_T.
\end{cases}
\]
Thus, for an arbitrary situation $\sigma$ for which $P(\sigma)$ is a finite play terminating at $v\in V_T$ and for a player $i\in I$ we have 
\[
\ell^i(\sigma)~=~ \sum_{(u,v)\in P(\sigma)} \ell^i(u,v) ~=~ \frac{\vert P(\sigma)\vert -1}{2\vert E\vert } + (M+\cL^i(v))
\]
implying that for all situations $\sigma$ with a finite play and for all players $i\in I$ we have 
\[
M+\cL^i(\sigma) ~\leq~ \ell^i(\sigma) ~<~ M+\cL^i(\sigma) ~+~ \frac12.
\]
Now assume that $\sigma$ is a terminal NE in $\Gamma'$ and let $\sigma'$ be obtained from $\sigma$ by one of the players, say $i\in I$ changing his strategy. If $\sigma'$ is infinite, then $\cL^i(\sigma')>\cL^i(\sigma)$. Otherwise, by the above inequalities we can write
\[
M+\cL^i(\sigma) \leq \ell^i(\sigma) \leq \ell^i(\sigma') < M+\cL^i(\sigma')+\frac12,
\]
where the second inequality follows from the fact that $\sigma$ is  NE in $\Gamma'$. Since the $\cL^i$ values are integers, $\cL^i(\sigma)\leq \cL^i(\sigma')$ follows, proving that $\sigma$ is also a NE in $\Gamma$.
\end{proof}

Note, that in the above construction, $\Gamma$ may have some NE that are not NE in $\Gamma'$.

Note also that $\Gamma$ and $\Gamma'$ in the above statement use the same underlying directed graph. Thus they are simultaneously edge-symmetric or non-edge-symmetric and we can derive the following claim:

\begin{corollary}\label{c1-SP->Terminal} 
	Edge-symmetric terminal games satisfying condition (CIW) have NE. Furthermore, there is a terminal NE whenever $V_T$  is reachable from  $v_0$. 
\end{corollary}

\begin{proof}
By Theorem \ref{t4-terminal=shortest-path}, an edge-symmetric terminal game satisfying condition (CIW) can be viewed as a positive shortest path game, and thus the claim follows by Theorem \ref{t1}.
\end{proof}

\bigskip

\section{Edge-Symmetric Terminal Games}

Let us consider an edge-symmetric terminal game $\Gamma=(G,v_0,\cL)$, and note that unlike in shortest path games, a loop (that is a move of the form $(u,u)$ for some position $u\in V$) may play a role in a NE, since the infinite play may not be the worse outcome for some players. Consequently, we have $u\in N^+(u)$ whenever $(u,u)\in E$.


Analogously to shortest path games, let us consider the family $\cQ$ of strongly connected components of the subgraphs induced by the subsets $V_i$, $i\in I$. 

We associate to $\Gamma$ its \emph{small} version $\Gamma'$ obtained by ``merging'' the strongly connected components $Q\in\cQ$ into single positions. More precisely, we introduce a new position $v_Q$ and set $i(v_Q)=i(Q)$  for all $Q\in\cQ$, and define a directed graph $G'$ on the set of positions $V'=V_T\cup \{v_Q\mid Q\in\cQ\}$. We define the edge set $E'$ of $G'$ by including edges $(v_Q,v_R)$ for $Q,R\in \cQ$, $Q\neq R$ if there are positions $u\in Q$ and $v\in R$ such that $(u,v)\in E$. We also include edge $(v_Q,w)$ for $Q\in\cQ$ and $w\in V_T$ if there is a position $u\in Q$ such that $(u,w)\in E$. 
Furthermore, we include a looping edge $(v_Q,v_Q)$ for all $Q\in\cQ$ with $\vert  Q\vert\geq 2$. Furthermore, if $Q=\{u\}$ for some $Q\in\cQ$ and $(u,u)\in E$, then we also include in $E'$ the loop $(v_Q,v_Q)$. Note that with the above definitions we avoided creating parallel edges, since they would be redundant in a terminal game. 
Finally, we define $v_0'=v_Q$ if $v_0\in Q$. Note that this small game $\Gamma'=(G',v_0',\cL)$ is again an edge-symmetric terminal game with the same set of players and cost function as $\Gamma$.

\begin{lemma}\label{l4}
	If the small game $\Gamma'$ has a NE then so does $\Gamma$.
\end{lemma}
\begin{proof}
Assume that $\sigma'\subseteq E'$ is a NE of $\Gamma'$. We associate to $\sigma'$ a situation $\sigma$ in $\Gamma$ as follows. 

For every position $v_Q\in V'$, $Q\in\cQ$ there exists a unique edge $(v_Q,x)\in\sigma'$. We associate to $(v_Q,x)$ edges of $E$ to be included in $\sigma$ in the following way:

If $x=v_R\neq v_Q$, then by the definition of the small game we have (at least one) corresponding edge $(u,v)\in E$ such that $u\in Q$, $v\in R$. Let us consider one of these $(u,v)$ edges, and declare position $u=u^Q$ the root of the strong component $Q\in \cQ$. Since $Q\in\cQ$ is a strong component, we have a directed tree $T^Q$ rooted at $u^Q\in Q$  such that from every vertex $u'\in Q$ we have a unique directed path from $u'$ to $u^Q$ via the edges of $T^Q$. 
Let us then include in $\sigma$ the edges of $T^Q$ and edge $(u,v)=(u^Q,v)$.

If $x=v_Q$ and $\vert Q\vert\geq 2$, then (by the definition of a component) $Q$ is a strongly connected subgraph with at least two vertices, $u,v\in Q$, $u\neq v$. Let us declare $u=u^Q$ the root of $Q$ and consider a directed tree $T^Q$ on the vertices of $Q$ such that from all other vertices of $Q$ there is a unique path to $u_Q$. Let us then include in $\sigma$ the edges of $T^Q$ and the edge $(u^Q,v)$.

Finally, if $x=v_Q$ and $Q=\{u\}$, then by the definition of the small game we must have $(u,u)\in E$, and  we include this loop in $\sigma$.

Note that we have $\cL^i(\sigma)=\cL^i(\sigma')$ for all $i\in I$. Furthermore, for all positions $v\in Q\in\cQ$ player $i(v)=i(Q)$ can reach the same set of terminals (and/or the infinite play) in $\Gamma$ and $\Gamma'$, assuming all other players keep their strategies. Thus, since $\sigma'$ is a NE in $\Gamma'$, situation $\sigma$ must also be a NE in $\Gamma$.
\end{proof}

\bigskip

\noindent{\textbf{Proof of Theorem \ref{t2}}:}
Assume that $\Gamma=(G, v_0, \cL)$ is an edge-symmetric terminal game. 
By Lemma \ref{l4} we can assume that $\vert Q\vert =1$ for all $Q\in \cQ$, or in other words that for all moves $(u,v)\in E$ we have $i(u)\neq i(v)$. 

Clearly, if $V_T$ is not reachable from $v_0$ then any situation is a NE.
Assume for the rest of our proof that $V_T$ is reachable from $v_0$.

For all positions $v\in V$ with $N^+(v)\cap V_T\neq \emptyset$ let us denote by $t(v)\in V_T$ the terminal that is a most preferred by player $i(v)$ in $N^+(v)\cap V_T$.

We consider the following cases:

\begin{description}
	\item[\textbf{Case 1:}] $N^+(v_0)\cap V_T=\emptyset$ and $\exists ~ v_1\in N^+(v_0)$ such that $N^+(v_1)\cap V_T=\emptyset$.\\
	In this case we can construct a NE $\sigma$ resulting in an infinite play. We include in $\sigma$ the moves $(x,v_0)$ for all $x\in N^+(v_0)$, $(v_0,v_1)$, and $(y,v_1)$ for all $y\in N^+(v_1)\setminus (N^+(v_0)\cup \{v_0\})$. For all other positions we include an arbitrary move. Here $P(\sigma)$ is the infinite play $v_0\to v_1\to v_0$, and neither $i(v_0)$ nor $i(v_1)$ can achieve a better outcome.
	Note that $v_0=v_1$ is possible in this case.
	
	\smallskip
	
	\item[\textbf{Case 2:}] $N^+(v_0)\cap V_T=\emptyset$ and $N^+(v)\cap V_T\neq\emptyset$ for all $v\in N^+(v_0)$.\\
	Let us define $v_1\in N^+(v_0)$ as a position such that $t(v_1)$ is one of the most preferred terminals for $i(v_0)$ in the subset $\{t(v)\mid v\in N^+(v_0)\}\subseteq V_T$.
	\begin{description}
		\item[\textbf{Subcase 2.1:}] Player $i(v_1)$ prefers $t(v_1)$ to an infinite play.\\
		We can construct a NE $\sigma$ in this case as follows. We include the move $(v_0,v_1)$, the moves $(x,v_1)$ for all $x\in N^+(v_1) \setminus V_T$, the moves $(v,t(v))$ for all $v\in N^+(v_0)\setminus N^+(v_1)$, and arbitrary moves for all other positions. Now the play $P(\sigma)$ is the path $v_0\to v_1\to t(v_1)$ yielding $\cL^i(\sigma)=\cL^i(t(v_1))$ for all players $i\in I$. Only two players are involved in the play, $i(v_0)$ and $i(v_1)$. If $i(v_1)$ deviates from $\sigma$ then he can get either another terminal in $N^+(v_1)\cap V_T$ or the infinite play, and neither one is better for him than $t(v_1)$. If $i(v_0)$ deviates then he can get one of the terminals $\{t(v)\mid v\in N^+(v_0)\}$, and by our choice, none of them is better for him than $t(v_1)$. 
		
		\smallskip
		
		\item[\textbf{Subcase 2.2:}] Player $i(v_1)$ prefers the infinite play at least as much as $t(v_1)$.\\
		We can construct a NE $\sigma$ in this case as follows. We include 
  the moves $(x,v_0)$ for all $x\in N^+(v_0)$, the moves $(y,v_1)$ for all $y\in N^+(v_1)\setminus V_T$, and an arbitrary move for all other positions. Now $P(\sigma)$ is the infinite play $v_0\to v_1\to v_0$. Player $i(v_0)$ can only deviate to another infinite play. Player $i(v_1)$ can deviate to either another infinite play, or to a terminal in $N^+(v_1)\cap V_T$, but those are not better for him then $t(v_1)$, which is not better in this case than an infinite play. 
	\end{description}
	\smallskip
	
	\item[\textbf{Case 3:}] $N^+(v_0)\cap V_T\neq\emptyset$.\\
	Let us now create a smaller subgame $\Gamma'=(G',v_0,\cL)$ by deleting from $\Gamma$ the moves $(v_0,w)$ for $w\in N^+(v_0)\cap V_T$. Since Case 1 or 2 holds for $\Gamma'$, we have a NE $\sigma'$ in $\Gamma'$ by the previous arguments. 
	\begin{description}
		\item[\textbf{Subcase 3.1:}] $\cL^{i(v_0)}(\sigma')\leq \cL^{i(v_0)}(t(v_0))$.\\
		In this case $\sigma'$ is also a NE in $\Gamma$, since only player $i(v_0)$ gained extra options, and all of those options would take him to a terminal in $N^+(v_0)\cap V_T$, which are assumed to be not better in this case for $i(v_0)$ than the outcome of $\sigma'$.
		
		\smallskip
		
		\item[\textbf{Subcase 3.2:}] $\cL^{i(v_0)}(\sigma')> \cL^{i(v_0)}(t(v_0))$.\\
		Let us modify $\sigma'$ by replacing the move from $v_0$ by $(v_0,t(v_0))$, and denote by $\sigma$ the obtained situation of $\Gamma$. We claim that $\sigma$ is a NE in $\Gamma$ in this case. This is because $P(\sigma)$ is the single move $v_0\to t(v_0)$, and any deviation from $\sigma$ by player $i(v_0)$ yields either another terminal in $N^+(v_0)\cap V_T$, which is not better than $t(v_0)$, or it yields $\sigma'$ or a situation obtainable by a deviation from $\sigma'$ in $\Gamma'$. Since $\sigma'$ is a NE in $\Gamma'$ none of these can give a better outcome to player $i(v_0)$ than $\sigma'$ which is not better than $t(v_0)$ by our assumptions in this case. 
	\end{description}
\end{description}
\qed

\bigskip

\section{Existence of UNE in edge-symmetric terminal games}\label{s-UNE}

In this section we prove Theorem \ref{t3} claiming that under conditions (TWO), (SYM), and (CIW) a terminal game has a UNE. 

We consider an edge-symmetric terminal game $\Gamma=(G,\cL)$ in which no initial position is fixed, and
assume that $I=\{1,2\}$. For a situation $\sigma$ and arbitrary position $v\in V$  we denote by $P(\sigma,v)$ the unique walk starting from position $v$ and following the moves in situation $\sigma$. Such a walk either terminates in a terminal position $w=w(\sigma,v)\in V_T$, or is infinite, cycling around a directed cycle infinitely many times. Accordingly, we define the effective cost for player $i\in I$ for a position $v\in V$ by
\[
\cL^i(\sigma,v) ~=~ \begin{cases}
	\cL^i(w(\sigma,v)) &\text{ if } P(\sigma,v) \text{ is terminating, and }\\
	0 & \text{ if } P(\sigma,v) \text{ is infinite.}
\end{cases}
\]
Note that the above definition and condition (CIW) imply that we have $\cL^i(w)<0$ for all players $i\in I$ and terminals $w\in V_T$. 

Given a player $i\in I$ we use $\sigma^{-i}$ to denote the strategy of the opponent. Thus, we can write $\sigma=(\sigma^{-i},\sigma_i)$ for an arbitrary situation and player $i\in I$, where $\sigma_i\in \Sigma_i$ is a strategy for player $i\in I$. Note that in this section we have $\vert I\vert=2$, and thus for a situation $\sigma=(\sigma_1,\sigma_2)$ we have $\sigma^{-1}=\sigma_2$ and $\sigma^{-2}=\sigma_1$. 

Given a player $i\in I$ and $\sigma^{-i}$ a strategy $\sigma_i\in \Sigma_i$ is called a \emph{uniform best response} of player $i$ to $\sigma^{-i}$ if the equality
\[
\cL^i((\sigma^{-i},\sigma_i),v) ~=~ \min_{\sigma_i'\in\Sigma_i}\cL^i((\sigma^{-i},\sigma_i'),v)
\]
holds for all positions $v\in V$. Note that a uniform best response always exists, can be determined efficiently, and may not be unique. 

We call a situation $\sigma=(\sigma_i\mid i\in I)$ a \emph{uniform Nash equilibrium} (or UNE, in short), if $\sigma_i$ is a uniform best response of player $i$ to $\sigma^{-i}$ for all players $i\in I$.

Given a situation $\sigma=(\sigma^{-i},\sigma_i)$ we say that strategy $\sigma_i'$ is a \emph{uniform best improvement} for player $i$ if $\sigma_i'$ is a uniform best response to $\sigma^{-i}$, different from $\sigma_i$, and for all positions $v\in V_i$ we have either $\sigma_i(v)=\sigma_i'(v)$ or $\cL^i((\sigma^{-i},\sigma_i'),v) < \cL^i((\sigma^{-i},\sigma_i),v)$. Note that every uniform best improvement is a uniform best response, but not necessarily the other way around. Furthermore,
if a situation is not a UNE, then at least one of the players have a uniform best improvement 
(and it can be determined efficiently). 
Let us also note that a uniform best improvement 
may not be unique in case players have ties over the set of terminals.

Our first step is to show that it is enough to prove Theorem \ref{t3} for simplified, special edge-symmetric terminal games. 
\begin{lemma}\label{l-UNE-0}
If $\Gamma=(G,\cL)$ is an edge-symmetric terminal game, then we can assume w.l.o.g. that edges are between different vertices controlled by different players, i.e.,
\begin{equation}\label{e-UNE-1}
	(u,v)\in E, ~u\neq v\not\in V_T ~~~\implies~~~ i(u)\neq i(v). 
\end{equation}
We can also assume that from each position we have at most one terminal move, i.e.,
\begin{equation}\label{e-UNE-1.5}
	\vert N^+(v)\cap V_T\vert  ~\leq~ 1 ~~~\text{ for all } v\in V.
\end{equation}
We can assume further that the terminals are reachable from each position $v\in V$, i.e.,
\begin{equation}\label{e-UNE-1.9}
	\text{ for all }~~ v\in V ~~\text{ there exist a }~~ v\to V_T ~~\text{ path.}
\end{equation}
\end{lemma}
\begin{proof}
For condition \eqref{e-UNE-1}
let us also note that Lemma \ref{l4} with the same proof works also for UNE (instead of NE). 

For the inequalities \eqref{e-UNE-1.5} note first that if $x,y\in N^+(v)\cap V_T$, $x\neq y$, and $\cL^{i(v)}(x)<\cL^{i(v)}(y)$ then the move $(v,y)$ can not belong to a UNE. Furthermore, if $\cL^{i(v)}(x)=\cL^{i(v)}(y)$ and $\sigma'$ is a UNE in $\Gamma'$ obtained from $\Gamma$ by deleting move $(v,y)$, then $\sigma'$ is also a UNE in $\Gamma$. This is because only player $i(v)$ has an option in $\Gamma$ that is not available in $\Gamma'$ (namely the move $(v,y)$) but that move cannot be part of a uniform best improvement of player $i(v)$ to $\sigma'^{-i(v)}$ since the move $(v,x)$ is available for player $i(v)$ in $\Gamma'$, $\sigma'$ is a UNE in $\Gamma'$, and $\cL^{i(v)}(x)=\cL^{i(v)}(y)$.

For property \eqref{e-UNE-1.9} let us consider the set of positions $U\subseteq V$ such that no directed path connects a vertex $u\in U$ to the set of terminals $V_T$. Then, assigning arbitrary moves to positions in $U$ does not change whether a situation is a UNE or not. 
Since identifying the set of positions from which the set of terminals is not reachable is a computationally easy task, and since those positions have no influence on the existence of a UNE, we can assume w.l.o.g. that we preprocess the input game, and delete all such positions, before any further analysis. 
\end{proof}

\begin{lemma}\label{l-UNE-1}
Assume that $\Gamma=(G,\cL)$ is an edge-symmetric terminal game, satisfying conditions \eqref{e-UNE-1} and $\sigma$ is a situation such that for one of the players, $i\in I$, strategy $\sigma_i$ is a uniform best response to $\sigma^{-i}$. Assume further that $u\in V_j$, $v\in V_i$ are distinct positions with $j\neq i$ that are connected by edges in $G$. Then we have
\[
\cL^i(\sigma,v) ~\leq~ \cL^i(\sigma,u).
\]
\end{lemma}
\begin{proof}
Let us note first that if $(v,u)$ belongs to $\sigma_i$ or $P(\sigma,u)$ contains position $v$, then we must have $\cL^i(\sigma,v) = \cL^i(\sigma,u)$ by the definition of a terminal game. 

Assume next that the move $(v,u)$ does not belong to $\sigma_i$ and position $v$ does not belong to $P(\sigma,u)$. Let us define a new strategy $\sigma_i'$ for player $i$ by
\[
\sigma_i'(w) ~=~ \begin{cases}
	\sigma_i(w) & \text{ for all } w\in V_i\setminus\{v\},\\
	u  & \text{ for } w=v.
\end{cases}
\]
With this definition we have
\[
\cL^i(\sigma,u) ~=~ \cL^i((\sigma^{-i},\sigma_i'),u) ~=~ \cL^i((\sigma^{-i},\sigma_i'),v) ~\geq~ \cL^i((\sigma^{-i},\sigma_i),v).
\]
Here the first two equalities follow by the definition of $\sigma_i'$ and the fact that position $v$ does not belong to $P(\sigma,u)$. The last inequality follows by the assumption that $\sigma_i$  is a uniform best response to $\sigma^{-i}$. 
\end{proof}

\bigskip
\begin{lemma}\label{l-UNE-2}
	Assume that $\Gamma=(G,\cL)$ is a terminal game satisfying condition \emph{(CIW)}, and $\sigma$ is a situation such that the plays $P(\sigma,v)$ are finite for all $v\in V$. Then, if $\sigma'$ is obtained from $\sigma$ by a uniform best improvement of one of the players, the plays $P(\sigma',v)$, $v\in V$ are also all finite.
\end{lemma}
\begin{proof}
This is an immediate consequence of condition (CIW) and the definition of uniform best improvement.
\end{proof}

Our proof of Theorem \ref{t3} is based on an iterative process the finiteness of which depends on a strictly monotone decreasing measure of progress. The proof of that monotonicity depends critically on the following definition.  

Let us assume, for the rest of this section, that by \eqref{e-UNE-1.5} of Lemma \ref{l-UNE-0} we have $\vert N^+(v)\cap V_T\vert  \leq 1$ for all $v\in V$. Let us then denote by $t(v)\in N^+(v)\cap V_T$ the unique terminal for vertices $v\in V$ with $N^+(v)\cap V_T\neq\emptyset$. 

Let us call a situation $\sigma$ \emph{$i$-basic} for a player $i\in I$ if for all positions $v\in V$ with $N^+(v)\cap V_T\neq \emptyset$ we either have $\sigma(v)=t(v)$ or $\cL^{i(v)}(\sigma',v)<\cL^{i(v)}(t(v))$ hold for all situations $\sigma'$ obtained from $\sigma$ by a uniform best improvement of player $i$. 
In other words, a situation is $i$-basic if a uniform best improvement by player $i$ provides every position that it controls and has an unused terminal move with a strictly better outcome than that terminal move would provide. 

\begin{lemma}\label{l-UNE-3}
Assume that $\Gamma=(G,\cL)$ is an edge-symmetric terminal game that satisfies the conditions in Lemma \ref{l-UNE-0}. Then we can find efficiently a situation $\sigma$ that is $i$-basic for all players $i\in I$, and for which the plays $P(\sigma,v)$ are finite for all $v\in V$.
\end{lemma}

\begin{proof}
We can construct a situation $\sigma$ satisfying the claimed properties by the following approach. Let us define $W=\{v\in V\mid N^+(v)\cap V_T\neq\emptyset\}$ and note that by our assumptions we have a terminal $t(v)\in V_T$ for all $v\in W$ such that $N^+(v)\cap V_T=\{t(v)\}$ holds. Let us now choose the moves $(v,t(v))$ for all positions $v\in W$, and color all positions in $W\cup V_T$ blue. In the sequel, while there exists an uncolored position $v\in V$ from which we can reach a blue position $u$ in one move, we choose the move $(v,u)$ and color $v$ blue. In a finite number of steps this procedure will stop, and all positions will be colored blue, since we assume that there is a finite directed path from all positions to the set of terminals. Thus we define in this way a situation $\sigma$ such that all plays $P(\sigma,v)$ are finite. 

To see that $\sigma$ is $i$-basic, for all players $i\in I$, let us consider one of the players $i$, and derive $\sigma'$ by applying a uniform best improvement of player $i$ to $\sigma$. Now, for a position $v\in W\setminus V_i$ we have $\sigma'(v)=\sigma(v)=t(v)$ by our definitions of $\sigma$ and $\sigma'$, while for a position $v\in W\cap V_i$ we have either $\sigma'(v)=\sigma(v)=t(v)$ or $\cL^i(\sigma',v)<\cL^i(\sigma,v)=\cL^i(t(v))$, by the definition of a uniform best improvement. 
\end{proof}

\bigskip
\begin{lemma}\label{l-UNE-4}
	Assume that $\Gamma=(G,\cL)$ is a $2$-person edge-symmetric terminal game satisfying conditions \emph{(CIW)}, conditions in Lemma \ref{l-UNE-0}, and that $\sigma$ is a $1$-basic situation such that all plays $P(\sigma,v)$ are finite. Assume further that
	$\sigma^1$ is obtained from $\sigma$ by a uniform best improvement of player $1$, and $\sigma^2$ is obtained from $\sigma^1$ by a uniform best improvement of player $2$.
	Then we have the inequalities 
	\[
	\cL^{i(v)}(\sigma^2,v) ~\leq~ \cL^{i(v)}(\sigma^1,v)
	\]
	for all positions $v\in V$. 
\end{lemma}
\begin{proof}
Since $\sigma^2$ is obtained from $\sigma^1$ by a uniform best improvement of player $2$, the claim follows for all $v\in V_2$ by the definition of a uniform best improvement. 

Let us next consider positions $v\in V_1$. If $P(\sigma^1,v)=P(\sigma^2,v)$, then we have the equality $\cL^{1}(\sigma^{2},v) = \cL^{1}(\sigma^1,v)$. 

Assume for the rest of our proof that $P(\sigma^2,v)\neq P(\sigma^1,v)$. Since $P(\sigma^2,v)$ is a finite play by Lemma \ref{l-UNE-2}, positions controlled by player $1$ and $2$ alternate along this finite path, due to our assumption \eqref{e-UNE-1}. Since $P(\sigma^2,v)\neq P(\sigma^1,v)$, we must have some positions $w\in V_2$ on the path $P(\sigma^2,v)$ such that $\sigma^2_2(w)\neq \sigma^1_2(w)$. Let us number all such positions as $w_1,..., w_k\in V_2$ for some $k\geq 1$, in the order we pass through them along the path $P(\sigma^2,v)$, when we start from $v$. Since $\sigma$ is $1$-basic, we cannot have $\sigma^2(w_k)\in V_T$, and thus we have $v_j=\sigma^2(w_j)\in V_1$ for all indices $j=1,...,k$. Now, let us observe that the moves $(v_j,w_j)$ exist, since this is an edge-symmetric terminal game, and thus they were considered in the uniform best improvement $\sigma^1$, for all $j=1,...,k$. Since they were not chosen, we must have the inequalities
\[
\cL^1(\sigma^1,v_j) ~\leq~ \cL^1(\sigma^1,w_j) 
\]
for all $j=1,...,k$ by Lemma \ref{l-UNE-1} applied for the moves $(v_j,w_j)$ with player $i=1$, $j=1,...,k$.
We also have the equalities
\[
\cL^1(\sigma^1,v_j)=\cL^1(\sigma^1,w_{j+1})
\]
for $j=1,...,k-1$, due to our selection of vertices $w_j$, $j=1,...,k$. These two groups of equations and inequalities imply that 
\[
\cL^1(\sigma^1,w_1) ~\geq~ \cL^1(\sigma^1,v_k).
\]
Since we also have $\cL^1(\sigma^2,v)=\cL^1(\sigma^2,v_k)=\cL^1(\sigma^1,v_k)$ and $\cL^1(\sigma^1,v)=\cL^1(\sigma^1,w_1)$, the stated inequalities follow.
\end{proof}

\bigskip
\begin{lemma}\label{l-UNE-5}
	Assume that $\Gamma=(G,\cL)$ is a $2$-person edge-symmetric terminal game satisfying conditions \emph{(CIW)}, conditions in Lemma \ref{l-UNE-0}, and that $\sigma$ is a $1$-basic situation in which all plays $P(\sigma,v)$ are finite. Assume further that
	$\sigma^1$ is obtained from $\sigma$ by a uniform best improvement of player $1$.
	Then situation $\sigma^1$ is $2$-basic.
\end{lemma}
\begin{proof}
Let us apply a uniform best improvement of player $2$ to $\sigma^1$, and denote the obtained situation by $\sigma^2$. 

Let us define $W=\{v\in V\mid N^+(v)\cap V_T\neq\emptyset\}$ and note that by Lemma \ref{l-UNE-0} we have a terminal $t(v)\in V_T$ for all $v\in W$ such that $N^+(v)\cap V_T=\{t(v)\}$ holds. 

Now, for a position $v\in W\cap V_2$ we have either $\sigma^2(v)=\sigma(v)$ and thus $\cL^2(\sigma^2,v)=\cL^2(\sigma,v)$, or we have $\sigma^2(v)\neq \sigma^1(v)=\sigma(v)$ and thus $\cL^2(\sigma^2,v) <\cL^2(\sigma^1,v)=\cL^2(\sigma,v)$ by the definition of a uniform best improvement. In both cases the claimed property follows by our assumption that $\sigma$ is $1$-basic. 

Finally, for positions $v\in W\cap V_1$ we have $\sigma^2(v)=\sigma^1(v)$. Thus, we either have $\sigma^2(v)=\sigma^1(v)=t(v)$, or $\sigma^1(v)\neq t(v)$, in which case we have $\cL_1(\sigma^1,v)<\cL^1(t(v))$ by the definition of a uniform best improvement. In this latter case we apply Lemma \ref{l-UNE-4} and conclude $\cL^1(\sigma^2,v)\leq\cL^1(\sigma^1,v)<\cL^1(t(v))$.
\end{proof}
\bigskip

\noindent{\textbf{Proof of Theorem \ref{t3}}:}
Let us first recall that the effective cost of an infinite play is defined as $0$, and thus condition (CIW) is equivalent with saying that $\cL^i(w)<0$ for all players $i\in I$ and terminals $w\in V_T$. We also assume that the simplifying conditions in Lemma \ref{l-UNE-0} hold.

Our proof idea is to show that if players alternate in making their uniform best improvements, then after a finite number of such iterations we arrive to a situation on which neither of the players can improve, showing that it is a UNE. While this idea would work starting with an arbitrary situation, our proof becomes simpler if we choose a special initial situation $\sigma^0=(\sigma^0_1,\sigma^0_2)$, in which all plays are finite and which is $1$-basic. According to Lemma \ref{l-UNE-3} such a situation exists and is easy to construct.

After this, players, starting with player $1$, alternate in making their uniform best improvements, as long as there is some change.
This creates a series of situations, $\sigma^j$, $j=1,2,...$. To be very precise, situation $\sigma^j$ for an odd $j\geq 1$ is obtained from $\sigma^{j-1}$ by a uniform best improvement of player $1$, while for an even $j\geq 2$ it is obtained from $\sigma^{j-1}$ by a uniform best improvement of player $2$. 

Note that by our definitions, if a player has a uniform best improvement, 
it actually improves strictly his local costs in at least 
some of the positions it controls, and degrades in neither. 
Thus the above process may terminate in a finite number of iterations, 
in which case the last situation is a UNE, by definition. 

On the other hand, a uniform best improvement by one of the players may actually increase some of the local costs for the other player, and thus the above procedure may start cycling through a number of situations, never terminating. To prove our theorem, we are going to show that such cycling cannot happen. To this end, we associate a quantity that can take only finitely many different values to every situation, and show that this  decreases strictly in every step in the above process. 

To a situation $\sigma$ let us associate for $i\in I$
\[
\nu^i(\sigma) ~=~ \sum_{v\in V_i} \cL^i(\sigma,v),
\]
and define the quantity associated to situations $\sigma$ as 
\[
\nu(\sigma) ~=~ \nu^1(\sigma)+\nu^2(\sigma).
\]
We claim that for all $j\geq 1$ we have
\begin{equation}\label{e-UNE-3}
\nu(\sigma^{j+1}) ~<~ \nu(\sigma^j).
\end{equation}

Note that by the definition of a uniform best improvement we have
\[
\begin{array}{rl}
	\nu^1(\sigma^{j+1}) &<~ \nu^1(\sigma^j) ~~~\text{ if $j$ is even, and } \\
	\nu^2(\sigma^{j+1}) &<~ \nu^2(\sigma^j) ~~~\text{ if $j$ is odd.} \\
\end{array}
\]
Thus, to prove our main claim \eqref{e-UNE-3}, and consequently the theorem, it is enough to show that for all $j\geq 1$ and positions $v\in V$ we have
\begin{equation}\label{e-UNE-5}
	\cL^{i(v)}(\sigma^{j+1},v) ~\leq~ \cL^{i(v)}(\sigma^j,v).
\end{equation}

Note that since $\sigma^0$ is $1$-basic, this claim follows by Lemma \ref{l-UNE-4} for $j=1$. 

By applying Lemma \ref{l-UNE-5}, recursively, we can conclude that $\sigma^j$ for an even $j\geq 2$ are all $1$-basic, and for an odd $j\geq 1$ are all $2$-basic. 

Note also that inequality \eqref{e-UNE-5} holds for positions $v\in V_1$ whenever $j$ is even, for $v\in V_2$ whenever $j$ is odd, simply by the definition of a uniform best improvement. For the remaining cases, we can prove \eqref{e-UNE-5} by induction on $j$, starting with $j=1$, and applying Lemma \ref{l-UNE-4}, with interchanging the players role's in it, alternately. 

The above arguments prove \eqref{e-UNE-5}, and thus \eqref{e-UNE-3} follows. Since $\nu(\sigma)$ can only take finitely many different values, this completes the proof of our theorem.
\qed

\begin{remark}
	Though our paper is not of algorithmic nature, we would like to remark that such a UNE for an edge-symmetric terminal game $\Gamma$ can be computed efficiently.
	
	First of all, computing a uniform best improvement is equivalent with solving a so-called reachability problem in $G$, and thus can be done in $O(\vert E\vert )$ time. 
	
	A starting strategy $\sigma^0$ can be computed in $O(\vert V\vert \vert E\vert )$ time, according to the procedure in Lemma \ref{l-UNE-3}. 
	
	According to our assumptions, as in Lemma \ref{l-UNE-0} and to condition (CIW), we can assume w.l.o.g. that
	\[
	-\vert V_T\vert  ~\leq~ \cL^i(w) ~<~ 0
	\]
	for all terminals $w\in V_T$ and players $i\in I$. Thus we have 
	\[
	-\vert V\vert \vert V_T\vert  ~\leq~ \nu(\sigma) <0
	\]
	for all situations $\sigma$. Thus we need only to consider at most $\vert V\vert \vert V_T\vert $ many uniform best improvements before termination, showing that our total complexity is $O(\vert V\vert \vert V_T\vert \vert E\vert )$.  
\end{remark}

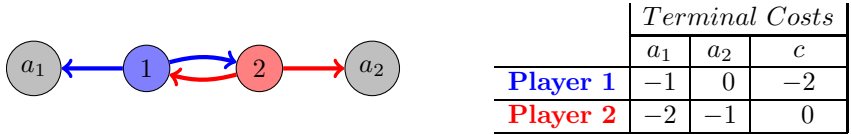
\begin{figure}[h]
	\begin{tikzpicture}[->,shorten >=0pt,auto,scale=1,node distance=1.5cm]
		
		\node[draw,circle,fill=gray!50] (a1) {$a_1$};
		\node[draw,circle,fill=blue!50] (u1) [right of=a1] {$1$};
		\node[draw,circle,fill=red!50] (u2) [right of=u1] {$2$};
		\node[draw,circle,fill=gray!50] (a2) [right of=u2] {$a_2$};
		
		\path[->,thin,gray]
		(u1) edge[bend left=15, ultra thick,blue] (u2)
		(u2) edge[bend left=15, ultra thick,red] (u1)
		(u1) edge[ultra thick,blue] (a1)
		(u2) edge[ultra thick,red] (a2);
		
		\node[right of=a2,right] {$
			\begin{array}{c|c|c|c|} 
				&\multicolumn{3}{c|}{Terminal~Costs}\\
				\cline{2-4}
				&a_1&a_2&c\\
				\hline
				\mathbf{\color{blue}Player~1}&-1&~~0&-2\\
				\hline
				\mathbf{\color{red}Player~2}&-2&-1&~~0\\
				\hline
			\end{array}
				$};
	\end{tikzpicture}
	\caption{Terminal game $\Gamma^2$ from \cite{BEGM12}. This example satisfies (TWO) and (SYM), but not (CIW) since \textbf{\color{blue}Player 1} prefers the infinite play, denoted by $c$ to both terminals $a_1$ and $a_2$. It is easy to verify that this game does not have a UNE. \label{fig-G2}}
\end{figure}

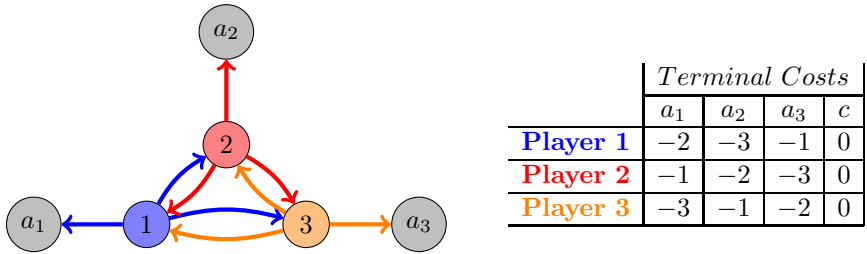
\begin{figure}[h]
	\begin{tikzpicture}[->,shorten >=0pt,auto,scale=1,node distance=1.5cm]
		
		\node[draw,circle,fill=gray!50] (a1) {$a_1$};
		\node[draw,circle,fill=blue!50] (u1) [right of=a1] {$1$};
		\node[draw,circle,fill=red!50] (u2) [above right of=u1] {$2$};
		\node[draw,circle,fill=gray!50] (a2) [above of=u2] {$a_2$};
		\node[draw,circle,fill=orange!50] (u3) [below right of=u2] {$3$};
		\node[draw,circle,fill=gray!50] (a3) [right of=u3] {$a_3$};
		
		\path[->,thin,gray]
(u1) edge[bend left=15, ultra thick,blue] (u2)
(u1) edge[bend left=15, ultra thick,blue] (u3)
(u2) edge[bend left=15, ultra thick,red] (u1)
(u3) edge[bend left=15, ultra thick,orange] (u1)
(u2) edge[bend left=15, ultra thick,red] (u3)
(u3) edge[bend left=15, ultra thick,orange] (u2)
(u1) edge[ultra thick,blue] (a1)
(u2) edge[ultra thick,red] (a2)
(u3) edge[ultra thick,orange] (a3);

		\node[above right of=a3,right] {$
			\begin{array}{c|c|c|c|c|} 
				&\multicolumn{4}{c|}{Terminal~Costs}\\
				\cline{2-5}
				&a_1&a_2&a_3&c\\
				\hline
				\mathbf{\color{blue}Player~1}&-2&-3&-1&0\\
				\hline
				\mathbf{\color{red}Player~2}&-1&-2&-3&0\\
				\hline
				\mathbf{\color{orange}Player~3}&-3&-1&-2&0\\
				\hline
			\end{array}
			$};
	\end{tikzpicture}
	\caption{A symmetrized variant of terminal game $\Gamma^3$ from \cite{BEGM12}. This example satisfies (CIW) and (SYM), but not (TWO). It is easy to verify that this game does not have a UNE. \label{fig-G3}}
\end{figure}

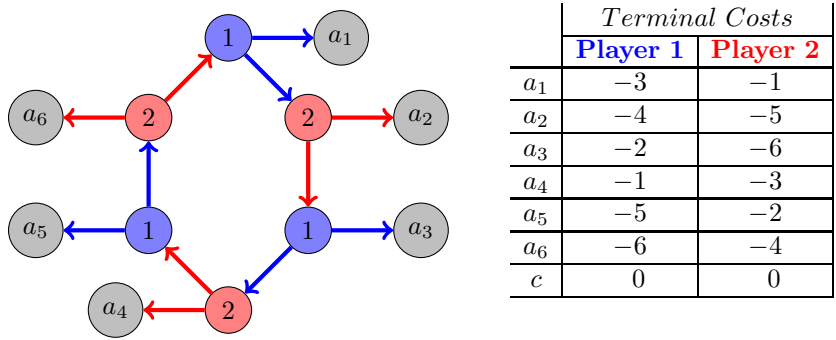
\begin{figure}
	\begin{tikzpicture}[->,shorten >=0pt,auto,scale=1,node distance=1.5cm]
		
		\node[draw,circle,fill=blue!50] (u1) {$1$};
		\node[draw,circle,fill=red!50] (u2) [below right of=u1] {$2$};
		\node[draw,circle,fill=blue!50] (u3) [below of=u2] {$1$};
		\node[draw,circle,fill=red!50] (u4) [below left of=u3] {$2$};
		\node[draw,circle,fill=blue!50] (u5) [above left of=u4] {$1$};
		\node[draw,circle,fill=red!50] (u6) [above of=u5] {$2$};
		\node[draw,circle,fill=gray!50] (a1) [right of=u1] {$a_1$};
			\node[draw,circle,fill=gray!50] (a2) [right of=u2] {$a_2$};
				\node[draw,circle,fill=gray!50] (a3) [right of=u3] {$a_3$};
					\node[draw,circle,fill=gray!50] (a4) [left of=u4] {$a_4$};
						\node[draw,circle,fill=gray!50] (a5) [left of=u5] {$a_5$};
							\node[draw,circle,fill=gray!50] (a6) [left of=u6] {$a_6$};

		\path[->,thin,gray]
		(u1) edge[ultra thick,blue] (u2)
		(u2) edge[ultra thick,red] (u3)
		(u3) edge[ultra thick,blue] (u4)
		(u4) edge[ultra thick,red] (u5)
		(u5) edge[ultra thick,blue] (u6)
		(u6) edge[ultra thick,red] (u1)
		(u1) edge[ultra thick,blue] (a1)
		(u2) edge[ultra thick,red] (a2)
		(u3) edge[ultra thick,blue] (a3)
		(u4) edge[ultra thick,red] (a4)
		(u5) edge[ultra thick,blue] (a5)
		(u6) edge[ultra thick,red] (a6);
		
		\node[above right of=a3,right] {$
			\begin{array}{c|c|c|} 
				&\multicolumn{2}{c|}{Terminal~Costs}\\
				\cline{2-3}
				&\mathbf{\color{blue}Player~1}&\mathbf{\color{red}Player~2}\\
				\hline
				a_1&-3&-1\\
				\hline
				a_2&-4&-5\\
				\hline
				a_3&-2&-6\\
				\hline
				a_4&-1&-3\\
				\hline
				a_5&-5&-2\\
				\hline
				a_6&-6&-4\\
				\hline
				c&~~0&~~0\\
				\hline
			\end{array}
			$};
	\end{tikzpicture}
	\caption{The terminal game $\Gamma^6$ from \cite{BEGM12}. This example satisfies (CIW) and (TWO), but not (SYM). We refer the reader to \cite{BEGM12} for a proof of the claim that this game does not have a UNE. \label{fig-G6}}
\end{figure}

\begin{remark}
	We also mention that this theorem is sharp in the sense that all three listed conditions, (TWO), (SYM), and (CIW) are necessary, as we can demonstrate this by the next three examples: 
	\begin{itemize}
		\item Consider $\Gamma^2$ from \cite{BEGM12} (see also Figure \ref{fig-G2}). This example satisfies (TWO) and (SYM), but not (CIW) and has no UNE. 
		\item Consider $\Gamma^3$ from \cite{BEGM12} and symmetrize it (see Figure \ref{fig-G3}). This example satisfies (SYM) and (CIW), but has three players and has no UNE.
		\item Consider finally $\Gamma^6$ from \cite{BEGM12} (see also Figure \ref{fig-G6}). This example satisfies (TWO) and (CIW), but not (SYM) and has no UNE. 
	\end{itemize}
	Most of these claims are very easy to check, based on the graphical description of these terminal games given in Figures \ref{fig-G2}, \ref{fig-G3}, and \ref{fig-G6}, except the last one, where the lack of a UNE is more complicated to check. We refer the reader to \cite{BEGM12} where this claim is demonstrated by describing the $8\times 8$ normal form of $\Gamma^6$.
\end{remark}

\bigskip

\section{NE-free Examples}\label{NEfree}

In this section we provide examples for edge-symmetric shortest path games that have no NE (and of course violate in some way the positivity condition). We also provide an example for a positive edge-symmetric shortest path game that has no UNE.

Let us recall (e.g., from \cite{BEGMO17,BEGM14}) that the effective cost of infinite plays can be defined in different ways, where differences are due to cycles in which the sum of edge lengths is zero (for some of the players). Accordingly, we provide two examples. 

\begin{figure}[ht]
	\centering
\scalebox{1}{
	\begin{tikzpicture}[->,shorten >=1pt,auto,node distance=5cm,semithick]
		
		\node[draw,circle,fill=red!50] (u1) [node distance=1.5cm] {$s$};
		\node[draw,circle,fill=blue!50] (u3) [above right of=u1] {$u$};
		\node[draw,circle,fill=blue!50] (u2) [below right of=u3] {$v$};
		\node[draw,fill=gray!50] (t) [right of=u2,node distance=1.5cm] {$t$};
		
		\path[->,thick,red]
		(u1) edge node[black,midway] {$\footnotesize\begin{array}{c}\br 1\\\bb 0\end{array}$} (u2)
		(u1) edge[bend left] node[black,sloped,above,midway] {$\footnotesize\begin{array}{c}\br -1\\\bb 2\end{array}$} (u3);
		\path[->,thick,blue]
		(u2) edge[bend left] node[black,midway] {$\footnotesize\begin{array}{c}\br 1\\\bb 1\end{array}$} (u1)
		(u2) edge node[black,sloped,above,midway] {$\footnotesize\begin{array}{c}\br 0\\\bb 0\end{array}$} (t)
		(u3) edge[bend left] node[black,sloped,above,midway] {$\footnotesize\begin{array}{c}\br 3\\\bb 2\end{array}$} (u2)
		(u2) edge[bend left] node[black,sloped,above,midway] {$\footnotesize\begin{array}{c}\br 0\\\bb 0\end{array}$} (u3)
		(u3) edge[bend left] node[black,sloped,above,midway] {$\footnotesize\begin{array}{c}\br 0\\\bb -1\end{array}$} (u1);

	\end{tikzpicture}
}
	\caption{A 2-player edge-symmetric non-zero sum shortest path game example with no NE. Edge lengths are arbitrary real numbers, and there is no zero-cycle.
		Player \textsc{\color{red}Red} controls the initial position $s$, and \textsc{\color{blue}Blue} controls positions $u$ and $v$. Both players are minimizing the length of the (infinite) path starting in the initial position $s$. Note that all stationary strategies yield a play that is ending in either a cycle yielding $\pm \infty$ as game value, or in a finite path terminating at $t$ and yielding finite values for the players. For example, the play $s\to v\to s$ has a value $-\infty$ for \textsc{\color{red}Red}, and $\infty$ for \textsc{\color{blue}Blue}.
		\label{fig1-pm}}
\end{figure}
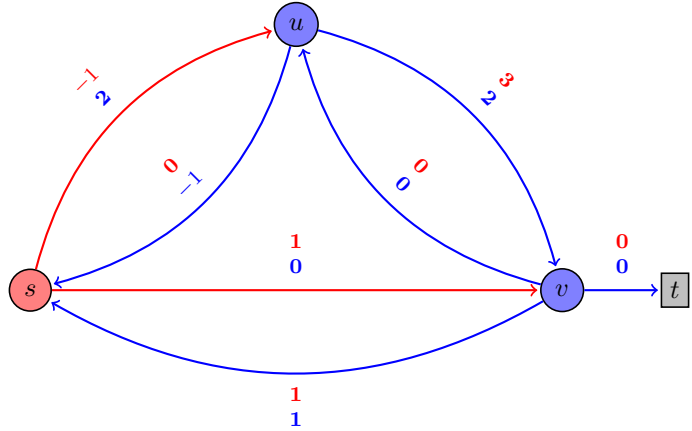

The first example shown in Figure \ref{fig1-pm} is a 2-person nonzero sum shortest path game on a symmetric digraph that has both positive and negative edge lengths and no cycle has length zero. The corresponding normal form is shown in Figure \ref{fig2-pm} proving that this example has no NE. 

\begin{figure}[ht]
	\centering
\scalebox{1}{
	\begin{tikzpicture}[->,shorten >=1pt,auto,node distance=5cm,semithick]
		
		\node[draw,circle,fill=red!50] (u1) [node distance=1.5cm] {$s$};
		\node[draw,circle,fill=blue!50] (u3) [above right of=u1] {$v$};
		\node[draw,circle,fill=blue!50] (u2) [below right of=u3] {$u$};
		\node[draw,fill=gray!50] (t) [right of=u2,node distance=1.5cm] {$t$};
		
		\path[->,thick,red]
		(u1) edge node[black,above,midway] {$\footnotesize\begin{array}{c}\br 1\\\bb 0\end{array}$} (u2)
		(u1) edge[bend left] node[black,sloped,above,midway] {$\footnotesize\begin{array}{c}\br 0\\\bb 1\end{array}$} (u3);
		\path[->,thick,blue]
		(u2) edge node[black,sloped,above,midway] {$\footnotesize\begin{array}{c}\br 0\\\bb 1\end{array}$} (t)
		(u3) edge[bend left] node[black,sloped,above,midway] {$\footnotesize\begin{array}{c}\br 2\\\bb 1\end{array}$} (u2)
		(u2) edge[bend left] node[black,sloped,above,midway] {$\footnotesize\begin{array}{c}\br 0\\\bb 0\end{array}$} (u3)
		(u3) edge[bend left] node[black,sloped,above,midway] {$\footnotesize\begin{array}{c}\br 0\\\bb 0\end{array}$} (u1)
		(u2) edge[bend left] node[black,sloped,above,midway] {$\footnotesize\begin{array}{c}\br 1\\\bb 1\end{array}$} (u1);
		
	\end{tikzpicture}
}
	\caption{A 2-player edge-symmetric non-zero sum shortest path game example with nonnegative edge lengths in which there is no NE. All edge lengths are nonnegative and some are equal to zero;  there are full zero cycles. Player \textsc{\color{red}Red} controls the initial position $s$, and \textsc{\color{blue}Blue} controls positions $u$ and $v$. Both players are minimizing the length of the (infinite) path starting at the initial position $s$. Note that all cycles are either positive (limit is $+\infty$), or have only arcs of length $0$. For instance, the play $s\to v\to s$ has value $0$ for \textsc{\color{red}Red} and $\infty$ for \textsc{\color{blue}Blue}.
		\label{fig1-p}
	}
\end{figure}
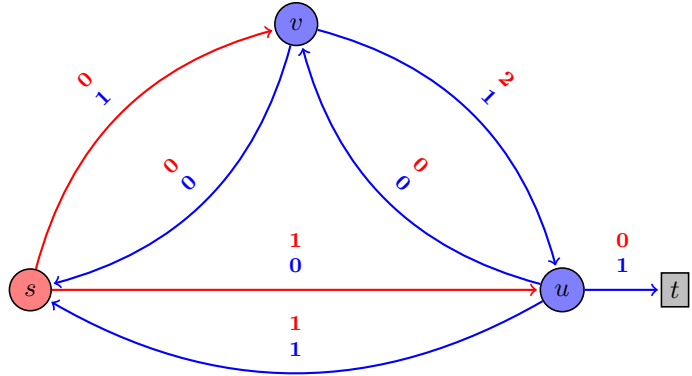

The second example shown in Figure \ref{fig1-p} is a 2-person nonzero sum shortest path game on a symmetric digraph that has nonnegative edge lengths but some of the cycles have zero length. In fact only cycles have zero length in which all edges have zero lengths. 
The corresponding normal form is shown in Figure \ref{fig4} proving that this example has no NE, either. 

Let us add that all meaningful definitions of the cost of an infinite 
play (that may lead to a NE) agree on the following facts (see \cite{BEGMO17,BEGM14}): 
if for a situation $\sigma$ the corresponding play $P=P(\sigma)$ ends in a cycle $C$ and $\sum_{e\in C} \ell^i(e)>0$, then $\ell^i(\sigma)=+\infty$, and if $\sum_{e\in C} \ell^i(e)<0$, then $\ell^i(\sigma)=-\infty$. Furthermore, if all edges $e\in C$ have $\ell^i(e)=0$, then $\ell^i(\sigma)=\sum_{e\in P\setminus C} \ell^i(e)$, or in other words, the effective cost of the play ending in cycle $C$ is the length of the path of $P$ leading to cycle $C$. In this sense, the above two examples show that positivity of the edge lengths in edge-symmetric shortest path games is "essential" to guarantee a NE.

\bigskip

\begin{figure}[ht]
	\centering
	\begin{tikzpicture}[->,shorten >=0pt,auto,scale=1,node distance=3.2cm]
		
		\node[draw,circle,fill=blue!50] (u1) {$1$};
		\node[draw,circle,fill=red!50] (u2) [below right of=u1] {$2$};
		\node[draw,circle,fill=blue!50] (u3) [below of=u2] {$1$};
		\node[draw,circle,fill=red!50] (u4) [below left of=u3] {$2$};
		\node[draw,circle,fill=blue!50] (u5) [above left of=u4] {$1$};
		\node[draw,circle,fill=red!50] (u6) [above of=u5] {$2$};
		\node[draw,circle,fill=gray!50] (a1) [right of=u1] {$a_1$};
		\node[draw,circle,fill=gray!50] (a2) [right of=u2] {$a_2$};
		\node[draw,circle,fill=gray!50] (a3) [right of=u3] {$a_3$};
		\node[draw,circle,fill=gray!50] (a4) [left of=u4] {$a_4$};
		\node[draw,circle,fill=gray!50] (a5) [left of=u5] {$a_5$};
		\node[draw,circle,fill=gray!50] (a6) [left of=u6] {$a_6$};

		\path[->,thin,gray]
		(u1) edge[ultra thick,bend left,blue] node[black,above,sloped,pos=0.6,scale=0.8] {$\footnotesize\begin{array}{c}
				\bb 0.01\\ \br 0.01
			\end{array}$} (u2)
		(u2) edge[ultra thick,bend left,red]  node[black,above,sloped,pos=0.5,scale=0.8] {$\footnotesize\begin{array}{c}
				\bb 0.01\\ \br 0.01
			\end{array}$} (u3)
		(u3) edge[ultra thick,bend left,blue]  node[black,above,sloped,pos=0.5,scale=0.8] {$\footnotesize\begin{array}{c}
				\bb 0.01\\ \br 0.01
			\end{array}$} (u4)
		(u4) edge[ultra thick,bend left,red]  node[black,above,sloped,pos=0.5,scale=0.8] {$\footnotesize\begin{array}{c}
				\bb 0.01\\ \br 0.01
			\end{array}$} (u5)
		(u5) edge[ultra thick,bend left,blue]  node[black,above,sloped,pos=0.5,scale=0.8] {$\footnotesize\begin{array}{c}
				\bb 0.01\\ \br 0.01
			\end{array}$} (u6)
		(u6) edge[ultra thick,bend left,red]  node[black,above,sloped,pos=0.5,scale=0.8] {$\footnotesize\begin{array}{c}
				\bb 0.01\\ \br 0.01
			\end{array}$} (u1)

(u1) edge[ultra thick,bend left,blue] node[black,above,sloped,pos=0.5,scale=0.8] {$\footnotesize\begin{array}{c}
		\bb 7\\ \br 7
	\end{array}$} (u6)
(u2) edge[ultra thick,bend left,red] node[black,above,sloped,pos=0.5,scale=0.8] {$\footnotesize\begin{array}{c}
		\bb 7\\ \br 7
	\end{array}$} (u1)
(u3) edge[ultra thick,bend left,blue] node[black,above,sloped,pos=0.5,scale=0.8] {$\footnotesize\begin{array}{c}
		\bb 7\\ \br 7
	\end{array}$} (u2)
(u4) edge[ultra thick,bend left,red] node[black,above,sloped,pos=0.5,scale=0.8] {$\footnotesize\begin{array}{c}
		\bb 7\\ \br 7
	\end{array}$} (u3)
(u5) edge[ultra thick,bend left,blue] node[black,above,sloped,pos=0.5,scale=0.8] {$\footnotesize\begin{array}{c}
		\bb 7\\ \br 7
	\end{array}$} (u4)
(u6) edge[ultra thick,bend left,red] node[black,above,sloped,pos=0.5,scale=0.8] {$\footnotesize\begin{array}{c}
		\bb 7\\ \br 7
	\end{array}$} (u5)

		(u1) edge[ultra thick,blue] node[black,above,sloped,pos=0.5,scale=0.8] {$\footnotesize\begin{array}{c}
				\bb 4\\ \br 6
			\end{array}$} (a1)
		(u2) edge[ultra thick,red] node[black,above,sloped,pos=0.5,scale=0.8] {$\footnotesize\begin{array}{c}
				\bb 3\\ \br 2
			\end{array}$} (a2)
		(u3) edge[ultra thick,blue] node[black,above,sloped,pos=0.5,scale=0.8] {$\footnotesize\begin{array}{c}
				\bb 5\\ \br 1
			\end{array}$} (a3)
		(u4) edge[ultra thick,red] node[black,above,sloped,pos=0.7,scale=0.8] {$\footnotesize\begin{array}{c}
				\bb 6\\ \br 4
			\end{array}$} (a4)
		(u5) edge[ultra thick,blue] node[black,above,sloped,pos=0.5,scale=0.8] {$\footnotesize\begin{array}{c}
				\bb 2\\ \br 5
			\end{array}$} (a5)
		(u6) edge[ultra thick,red] node[black,above,sloped,pos=0.5,scale=0.8] {$\footnotesize\begin{array}{c}
				\bb 1\\ \br 3
			\end{array}$} (a6);
		
	\end{tikzpicture}
	\caption{A $2$-person positive edge-symmetric shortest path game $\Gamma^{6s}$ derived from the terminal game $\Gamma^6$ in \cite{BEGM12}.\label{fig-G6s}}
\end{figure}
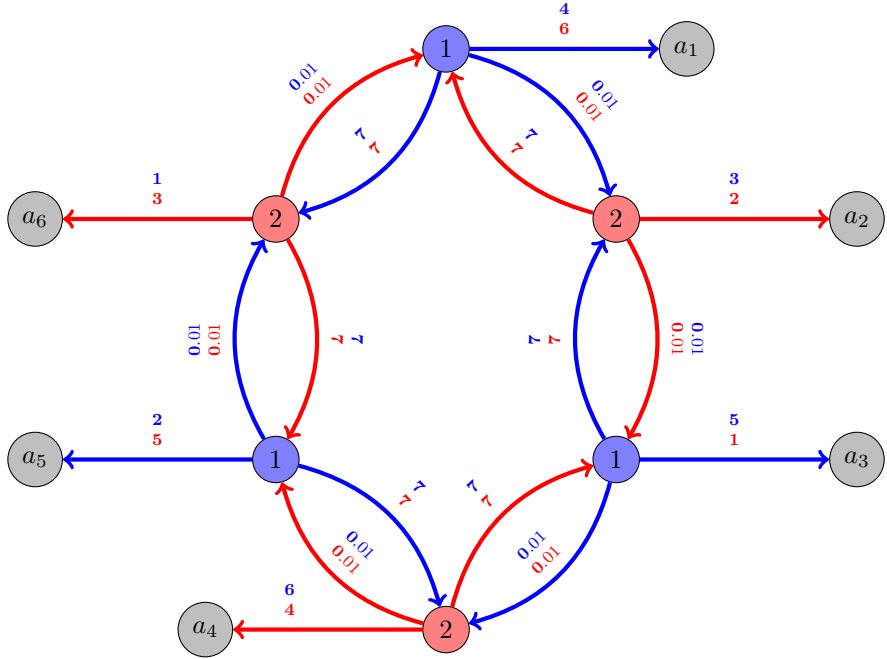

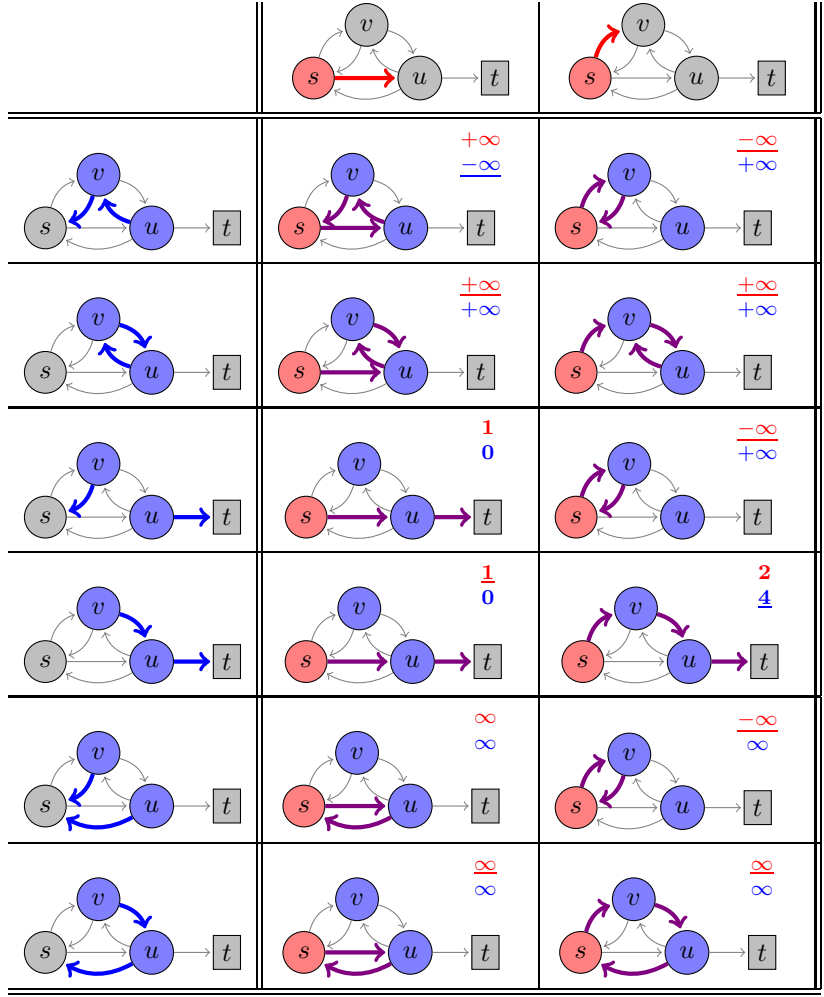
\begin{figure}[ht]
\begin{tabular}{ c||c|c|| } 

 &\begin{tikzpicture}[->,shorten >=1pt,auto,thin,scale=0.25]
 
 \node[draw,circle,fill=red!50] (u1) {$s$};
 \node[draw,circle,fill=gray!50] (u3) [above right of=u1] {$v$};
 \node[draw,circle,fill=gray!50] (u2) [below right of=u3] {$u$};
 \node[draw,fill=gray!50] (t) [right of=u2] {$\tiny t$};
 
 \path[->,thin,gray]
   (u1) edge (u2)
   (u2) edge[bend left] (u1)
   (u1) edge[bend left] (u3)
   (u2) edge[bend left] (u3)
   (u2) edge (t)
   (u3) edge[bend left] (u2)
   (u3) edge[bend left] (u1);
   
\path[->,line width=1.6pt,red] (u1) edge (u2);
  
 \end{tikzpicture}
  &\begin{tikzpicture}[->,shorten >=1pt,auto,thin,scale=0.25]
  
  \node[draw,circle,fill=red!50] (u1) {$s$};
  \node[draw,circle,fill=gray!50] (u3) [above right of=u1] {$v$};
  \node[draw,circle,fill=gray!50] (u2) [below right of=u3] {$u$};
  \node[draw,fill=gray!50] (t) [right of=u2] {$\tiny t$};
  
   \path[->,thin,gray]
     (u1) edge (u2)
     (u2) edge[bend left] (u1)
     (u1) edge[bend left] (u3)
     (u2) edge[bend left] (u3)
     (u2) edge (t)
     (u3) edge[bend left] (u2)
     (u3) edge[bend left] (u1);
     
\path[->,line width=1.6pt,red] (u1) edge[bend left] (u3);
    
  \end{tikzpicture}
   \\ 
   \hline
   \hline
 \begin{tikzpicture}[->,shorten >=1pt,auto,thin,scale=0.25]
 
 \node[draw,circle,fill=gray!50] (u1) {$s$};
 \node[draw,circle,fill=blue!50] (u3) [above right of=u1] {$v$};
 \node[draw,circle,fill=blue!50] (u2) [below right of=u3] {$u$};
 \node[draw,fill=gray!50] (t) [right of=u2] {$\tiny t$};
 
  \path[->,thin,gray]
    (u1) edge (u2)
    (u2) edge[bend left] (u1)
    (u1) edge[bend left] (u3)
    (u2) edge[bend left] (u3)
    (u2) edge (t)
    (u3) edge[bend left] (u2)
    (u3) edge[bend left] (u1);
    
\path[->,line width=1.6pt,blue]
 (u3) edge[bend left] (u1)
 (u2) edge[bend left] (u3);
   
 \end{tikzpicture}
 & \begin{tikzpicture}[->,shorten >=1pt,auto,thin,scale=0.25]
 
 \node[draw,circle,fill=red!50] (u1) {$s$};
 \node[draw,circle,fill=blue!50] (u3) [above right of=u1] {$v$};
 \node[draw,circle,fill=blue!50] (u2) [below right of=u3] {$u$};
 \node[draw,fill=gray!50] (t) [right of=u2] {$\tiny t$};
 
  \path[->,thin,gray]
    (u1) edge (u2)
    (u2) edge[bend left] (u1)
    (u1) edge[bend left] (u3)
    (u2) edge[bend left] (u3)
    (u2) edge (t)
    (u3) edge[bend left] (u2)
    (u3) edge[bend left] (u1);
    
\path[->,line width=1.6pt,red!50!blue]
 (u1) edge (u2)
 (u2) edge[bend left] (u3)
 (u3) edge[bend left] (u1);

 \node[above of=t,node distance=1cm] {$\footnotesize\begin{array}{c}\br +\infty\\\underline{\bb -\infty}\end{array}$};
   
 \end{tikzpicture}
 &  \begin{tikzpicture}[->,shorten >=1pt,auto,thin,scale=0.25]
 
 \node[draw,circle,fill=red!50] (u1) {$s$};
 \node[draw,circle,fill=blue!50] (u3) [above right of=u1] {$v$};
 \node[draw,circle,fill=blue!50] (u2) [below right of=u3] {$u$};
 \node[draw,fill=gray!50] (t) [right of=u2] {$\tiny t$};
 
  \path[->,thin,gray]
    (u1) edge (u2)
    (u2) edge[bend left] (u1)
    (u1) edge[bend left] (u3)
    (u2) edge[bend left] (u3)
    (u2) edge (t)
    (u3) edge[bend left] (u2)
    (u3) edge[bend left] (u1);
 
\path[->,line width=1.6pt,red!50!blue]
 (u1) edge[bend left] (u3)
 (u3) edge[bend left] (u1);
   
 \node[above of=t,node distance=1cm] {$\footnotesize\begin{array}{c}\underline{\br -\infty}\\\bb +\infty\end{array}$};
   
 \end{tikzpicture}
 \\ 
\hline
 \begin{tikzpicture}[->,shorten >=1pt,auto,thin,scale=0.25]

 \node[draw,circle,fill=gray!50] (u1) {$s$};
 \node[draw,circle,fill=blue!50] (u3) [above right of=u1] {$v$};
 \node[draw,circle,fill=blue!50] (u2) [below right of=u3] {$u$};
 \node[draw,fill=gray!50] (t) [right of=u2] {$\tiny t$};
 
  \path[->,thin,gray]
    (u1) edge (u2)
    (u2) edge[bend left] (u1)
    (u1) edge[bend left] (u3)
    (u2) edge[bend left] (u3)
    (u2) edge (t)
    (u3) edge[bend left] (u2)
    (u3) edge[bend left] (u1);
    
\path[->,line width=1.6pt,blue]
 (u3) edge[bend left] (u2)
 (u2) edge[bend left] (u3);
   
 \end{tikzpicture}
 & \begin{tikzpicture}[->,shorten >=1pt,auto,thin,scale=0.25]
 
 \node[draw,circle,fill=red!50] (u1) {$s$};
 \node[draw,circle,fill=blue!50] (u3) [above right of=u1] {$v$};
 \node[draw,circle,fill=blue!50] (u2) [below right of=u3] {$u$};
 \node[draw,fill=gray!50] (t) [right of=u2] {$\tiny t$};
 
 \path[->,thin,gray]
   (u1) edge (u2)
   (u2) edge[bend left] (u1)
   (u1) edge[bend left] (u3)
   (u2) edge[bend left] (u3)
   (u2) edge (t)
   (u3) edge[bend left] (u2)
   (u3) edge[bend left] (u1);
   
\path[->,line width=1.6pt,red!50!blue]
 (u1) edge (u2)
 (u2) edge[bend left] (u3)
 (u3) edge[bend left] (u2);

 \node[above of=t,node distance=1cm] {$\footnotesize\begin{array}{c}\underline{\br +\infty}\\\bb +\infty\end{array}$};
   
 \end{tikzpicture}
 &  \begin{tikzpicture}[->,shorten >=1pt,auto,thin,scale=0.25]
 
 \node[draw,circle,fill=red!50] (u1) {$s$};
 \node[draw,circle,fill=blue!50] (u3) [above right of=u1] {$v$};
 \node[draw,circle,fill=blue!50] (u2) [below right of=u3] {$u$};
 \node[draw,fill=gray!50] (t) [right of=u2] {$\tiny t$};
 
  \path[->,thin,gray]
    (u1) edge (u2)
    (u2) edge[bend left] (u1)
    (u1) edge[bend left] (u3)
    (u2) edge[bend left] (u3)
    (u2) edge (t)
    (u3) edge[bend left] (u2)
    (u3) edge[bend left] (u1);
 
\path[->,line width=1.6pt,red!50!blue]
 (u1) edge[bend left] (u3)
 (u3) edge[bend left] (u2)
 (u2) edge[bend left] (u3);
   
 \node[above of=t,node distance=1cm] {$\footnotesize\begin{array}{c}\underline{\br +\infty}\\\bb +\infty\end{array}$};
   
 \end{tikzpicture}
 \\ 
\hline
 \begin{tikzpicture}[->,shorten >=1pt,auto,thin,scale=0.25]
 
 \node[draw,circle,fill=gray!50] (u1) {$s$};
 \node[draw,circle,fill=blue!50] (u3) [above right of=u1] {$v$};
 \node[draw,circle,fill=blue!50] (u2) [below right of=u3] {$u$};
 \node[draw,fill=gray!50] (t) [right of=u2] {$\tiny t$};
 
  \path[->,thin,gray]
    (u1) edge (u2)
    (u2) edge[bend left] (u1)
    (u1) edge[bend left] (u3)
    (u2) edge[bend left] (u3)
    (u2) edge (t)
    (u3) edge[bend left] (u2)
    (u3) edge[bend left] (u1);
    
\path[->,line width=1.6pt,blue]
 (u3) edge[bend left] (u1)
 (u2) edge (t);
   
 \end{tikzpicture}
 & \begin{tikzpicture}[->,shorten >=1pt,auto,thin,scale=0.25]
 
 \node[draw,circle,fill=red!50] (u1) {$s$};
 \node[draw,circle,fill=blue!50] (u3) [above right of=u1] {$v$};
 \node[draw,circle,fill=blue!50] (u2) [below right of=u3] {$u$};
 \node[draw,fill=gray!50] (t) [right of=u2] {$\tiny t$};
 
 \path[->,thin,gray]
   (u1) edge (u2)
   (u2) edge[bend left] (u1)
   (u1) edge[bend left] (u3)
   (u2) edge[bend left] (u3)
   (u2) edge (t)
   (u3) edge[bend left] (u2)
   (u3) edge[bend left] (u1);
   
\path[->,line width=1.6pt,red!50!blue]
 (u1) edge (u2)
 (u2) edge (t);

 \node[above of=t,node distance=1cm] {$\footnotesize\begin{array}{c}\br 1\\\bb 0\end{array}$};
   
 \end{tikzpicture}
 &  \begin{tikzpicture}[->,shorten >=1pt,auto,thin,scale=0.25]
 
 \node[draw,circle,fill=red!50] (u1) {$s$};
 \node[draw,circle,fill=blue!50] (u3) [above right of=u1] {$v$};
 \node[draw,circle,fill=blue!50] (u2) [below right of=u3] {$u$};
 \node[draw,fill=gray!50] (t) [right of=u2] {$\tiny t$};
 
 \path[->,thin,gray]
   (u1) edge (u2)
   (u2) edge[bend left] (u1)
   (u1) edge[bend left] (u3)
   (u2) edge[bend left] (u3)
   (u2) edge (t)
   (u3) edge[bend left] (u2)
   (u3) edge[bend left] (u1);

\path[->,line width=1.6pt,red!50!blue]
 (u1) edge[bend left] (u3)
 (u3) edge[bend left] (u1);
   
 \node[above of=t,node distance=1cm] {$\footnotesize\begin{array}{c}\underline{\br -\infty}\\\bb +\infty\end{array}$};
   
 \end{tikzpicture}
 \\ 
\hline
 \begin{tikzpicture}[->,shorten >=1pt,auto,thin,scale=0.25]
 
 \node[draw,circle,fill=gray!50] (u1) {$s$};
 \node[draw,circle,fill=blue!50] (u3) [above right of=u1] {$v$};
 \node[draw,circle,fill=blue!50] (u2) [below right of=u3] {$u$};
 \node[draw,fill=gray!50] (t) [right of=u2] {$\tiny t$};
 
 \path[->,thin,gray]
   (u1) edge (u2)
   (u2) edge[bend left] (u1)
   (u1) edge[bend left] (u3)
   (u2) edge[bend left] (u3)
   (u2) edge (t)
   (u3) edge[bend left] (u2)
   (u3) edge[bend left] (u1);
  
\path[->,line width=1.6pt,blue]
 (u3) edge[bend left] (u2)
 (u2) edge (t);
   
 \end{tikzpicture}
 & \begin{tikzpicture}[->,shorten >=1pt,auto,thin,scale=0.25]
 
 \node[draw,circle,fill=red!50] (u1) {$s$};
 \node[draw,circle,fill=blue!50] (u3) [above right of=u1] {$v$};
 \node[draw,circle,fill=blue!50] (u2) [below right of=u3] {$u$};
 \node[draw,fill=gray!50] (t) [right of=u2] {$\tiny t$};
 
 \path[->,thin,gray]
   (u1) edge (u2)
   (u2) edge[bend left] (u1)
   (u1) edge[bend left] (u3)
   (u2) edge[bend left] (u3)
   (u2) edge (t)
   (u3) edge[bend left] (u2)
   (u3) edge[bend left] (u1);
   
\path[->,line width=1.6pt,red!50!blue]
 (u1) edge (u2)
 (u2) edge (t);

 \node[above of=t,node distance=1cm] {$\footnotesize\begin{array}{c}\underline{\br 1}\\\bb 0\end{array}$};
   
 \end{tikzpicture}
 &  \begin{tikzpicture}[->,shorten >=1pt,auto,thin,scale=0.25]
 
 \node[draw,circle,fill=red!50] (u1) {$s$};
 \node[draw,circle,fill=blue!50] (u3) [above right of=u1] {$v$};
 \node[draw,circle,fill=blue!50] (u2) [below right of=u3] {$u$};
 \node[draw,fill=gray!50] (t) [right of=u2] {$\tiny t$};
 
 \path[->,thin,gray]
   (u1) edge (u2)
   (u2) edge[bend left] (u1)
   (u1) edge[bend left] (u3)
   (u2) edge[bend left] (u3)
   (u2) edge (t)
   (u3) edge[bend left] (u2)
   (u3) edge[bend left] (u1);

\path[->,line width=1.6pt,red!50!blue]
 (u1) edge[bend left] (u3)
 (u3) edge[bend left] (u2)
 (u2) edge (t);
   
 \node[above of=t,node distance=1cm] {$\footnotesize\begin{array}{c}\br 2\\\underline{\bb 4}\end{array}$};
   
 \end{tikzpicture}
 \\ 
\hline
\begin{tikzpicture}[->,shorten >=1pt,auto,thin,scale=0.25]
  
  \node[draw,circle,fill=gray!50] (u1) {$s$};
  \node[draw,circle,fill=blue!50] (u3) [above right of=u1] {$v$};
  \node[draw,circle,fill=blue!50] (u2) [below right of=u3] {$u$};
  \node[draw,fill=gray!50] (t) [right of=u2] {$\tiny t$};
  
  \path[->,thin,gray]
    (u1) edge (u2)
    (u2) edge[bend left] (u1)
    (u1) edge[bend left] (u3)
    (u2) edge[bend left] (u3)
    (u2) edge (t)
    (u3) edge[bend left] (u2)
    (u3) edge[bend left] (u1);
   
 \path[->,line width=1.6pt,blue]
  (u3) edge[bend left] (u1)
  (u2) edge[bend left] (u1);
    
  \end{tikzpicture}
  & \begin{tikzpicture}[->,shorten >=1pt,auto,thin,scale=0.25]
  
  \node[draw,circle,fill=red!50] (u1) {$s$};
  \node[draw,circle,fill=blue!50] (u3) [above right of=u1] {$v$};
  \node[draw,circle,fill=blue!50] (u2) [below right of=u3] {$u$};
  \node[draw,fill=gray!50] (t) [right of=u2] {$\tiny t$};
  
  \path[->,thin,gray]
    (u1) edge (u2)
    (u2) edge[bend left] (u1)
    (u1) edge[bend left] (u3)
    (u2) edge[bend left] (u3)
    (u2) edge (t)
    (u3) edge[bend left] (u2)
    (u3) edge[bend left] (u1);
    
 \path[->,line width=1.6pt,red!50!blue]
  (u1) edge (u2)
  (u2) edge[bend left] (u1);

  \node[above of=t,node distance=1cm] {$\footnotesize\begin{array}{c}\br \infty\\\bb \infty\end{array}$};
    
  \end{tikzpicture}
  &  \begin{tikzpicture}[->,shorten >=1pt,auto,thin,scale=0.25]
  
  \node[draw,circle,fill=red!50] (u1) {$s$};
  \node[draw,circle,fill=blue!50] (u3) [above right of=u1] {$v$};
  \node[draw,circle,fill=blue!50] (u2) [below right of=u3] {$u$};
  \node[draw,fill=gray!50] (t) [right of=u2] {$\tiny t$};
  
  \path[->,thin,gray]
    (u1) edge (u2)
    (u2) edge[bend left] (u1)
    (u1) edge[bend left] (u3)
    (u2) edge[bend left] (u3)
    (u2) edge (t)
    (u3) edge[bend left] (u2)
    (u3) edge[bend left] (u1);
 
 \path[->,line width=1.6pt,red!50!blue]
  (u1) edge[bend left] (u3)
  (u3) edge[bend left] (u1);
    
  \node[above of=t,node distance=1cm] {$\footnotesize\begin{array}{c}\underline{\br -\infty}\\\bb \infty\end{array}$};
    
  \end{tikzpicture}
  \\ 
\hline
\begin{tikzpicture}[->,shorten >=1pt,auto,thin,scale=0.25]
 
 \node[draw,circle,fill=gray!50] (u1) {$s$};
 \node[draw,circle,fill=blue!50] (u3) [above right of=u1] {$v$};
 \node[draw,circle,fill=blue!50] (u2) [below right of=u3] {$u$};
 \node[draw,fill=gray!50] (t) [right of=u2] {$\tiny t$};
 
 \path[->,thin,gray]
   (u1) edge (u2)
   (u2) edge[bend left] (u1)
   (u1) edge[bend left] (u3)
   (u2) edge[bend left] (u3)
   (u2) edge (t)
   (u3) edge[bend left] (u2)
   (u3) edge[bend left] (u1);
  
\path[->,line width=1.6pt,blue]
 (u3) edge[bend left] (u2)
 (u2) edge[bend left] (u1);
   
 \end{tikzpicture}
 & \begin{tikzpicture}[->,shorten >=1pt,auto,thin,scale=0.25]
 
 \node[draw,circle,fill=red!50] (u1) {$s$};
 \node[draw,circle,fill=blue!50] (u3) [above right of=u1] {$v$};
 \node[draw,circle,fill=blue!50] (u2) [below right of=u3] {$u$};
 \node[draw,fill=gray!50] (t) [right of=u2] {$\tiny t$};
 
 \path[->,thin,gray]
   (u1) edge (u2)
   (u2) edge[bend left] (u1)
   (u1) edge[bend left] (u3)
   (u2) edge[bend left] (u3)
   (u2) edge (t)
   (u3) edge[bend left] (u2)
   (u3) edge[bend left] (u1);
   
\path[->,line width=1.6pt,red!50!blue]
 (u1) edge (u2)
 (u2) edge[bend left] (u1);

 \node[above of=t,node distance=1cm] {$\footnotesize\begin{array}{c}\underline{\br \infty}\\\bb \infty\end{array}$};
   
 \end{tikzpicture}
 &  \begin{tikzpicture}[->,shorten >=1pt,auto,thin,scale=0.25]
 
 \node[draw,circle,fill=red!50] (u1) {$s$};
 \node[draw,circle,fill=blue!50] (u3) [above right of=u1] {$v$};
 \node[draw,circle,fill=blue!50] (u2) [below right of=u3] {$u$};
 \node[draw,fill=gray!50] (t) [right of=u2] {$\tiny t$};
 
 \path[->,thin,gray]
   (u1) edge (u2)
   (u2) edge[bend left] (u1)
   (u1) edge[bend left] (u3)
   (u2) edge[bend left] (u3)
   (u2) edge (t)
   (u3) edge[bend left] (u2)
   (u3) edge[bend left] (u1);

\path[->,line width=1.6pt,red!50!blue]
 (u1) edge[bend left] (u3)
 (u3) edge[bend left] (u2)
 (u2) edge[bend left] (u1);
   
 \node[above of=t,node distance=1cm] {$\footnotesize\begin{array}{c}\underline{\br \infty}\\{\bb \infty}\end{array}$};
   
 \end{tikzpicture}
 \\ 
\hline
\hline
\end{tabular}
\caption{Normal form of Example given in Figure \ref{fig1-pm}. The columns represent the two possible strategies of player \textsc{\color{red}Red} and the rows represent the six possible strategies of player \textsc{\color{blue}Blue}. Both players are minimizing their respective costs, indicated in the upper right corner in each cell. Row minimizing and column minimizing costs are underlined. Since no situation is simultaneously row and column minimizing, this example has no NE.
\label{fig2-pm}}
\end{figure}

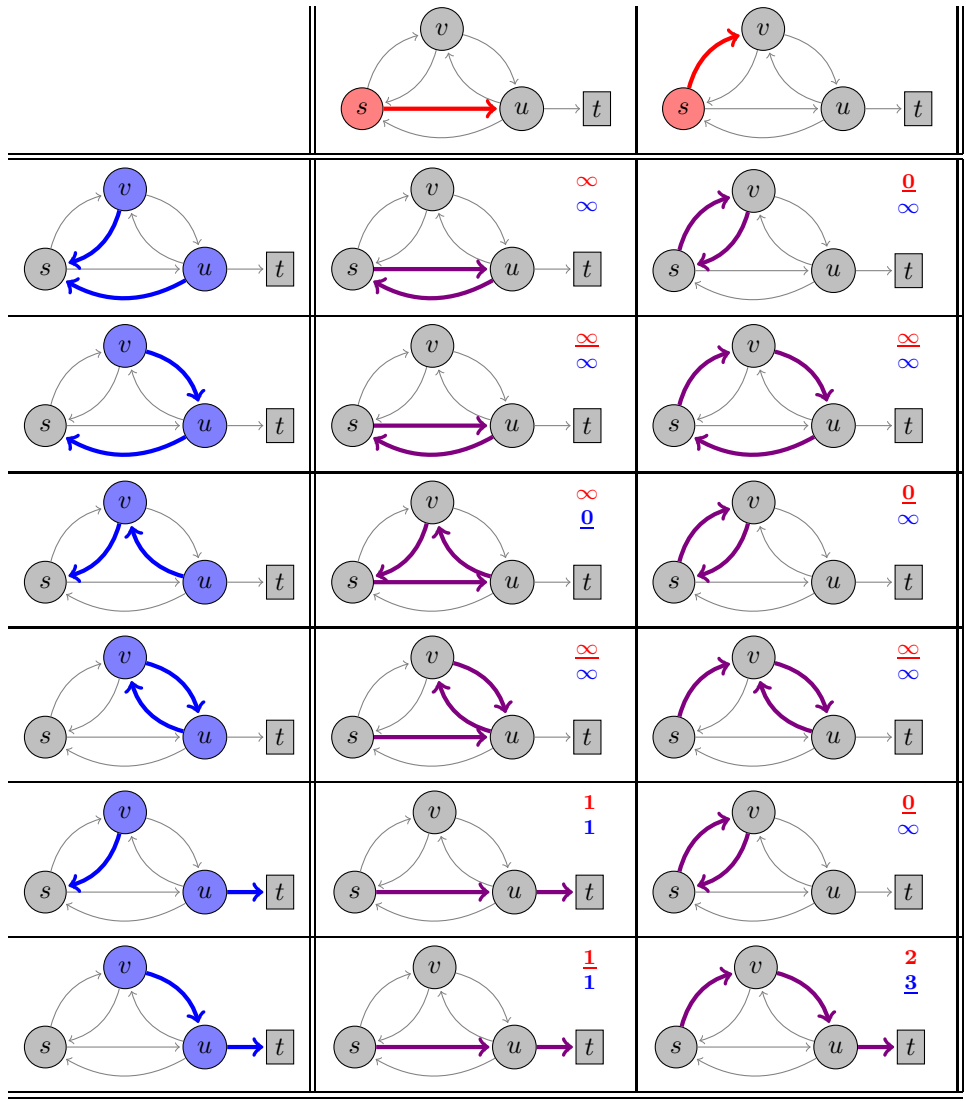
\begin{figure}[ht]
\begin{tabular}{ c||c|c|| } 

 &
\begin{tikzpicture}[->,shorten >=1pt,auto,node distance=1.5cm,thin]
 \node[draw,circle,fill=red!50] (u1) {$\tiny s$};
 \node[draw,circle,fill=gray!50] (u3) [above right of=u1] {$\tiny v$};
 \node[draw,circle,fill=gray!50] (u2) [below right of=u3] {$\tiny u$};
 \node[draw,fill=gray!50] (t) [right of=u2,node distance=1cm] {$\tiny t$};
 
 \path[->,thin,gray]
   (u1) edge (u2)
   (u1) edge[bend left] (u3)
   (u2) edge[bend left] (u3)
   (u2) edge (t)
   (u3) edge[bend left] (u2)
   (u3) edge[bend left] (u1)
   (u2) edge[bend left] (u1);
\path[->,line width=1.6pt,red] (u1) edge (u2);
\end{tikzpicture}
 &
\begin{tikzpicture}[->,shorten >=1pt,auto,node distance=1.5cm,thin]
  \node[draw,circle,fill=red!50] (u1) {$\tiny s$};
  \node[draw,circle,fill=gray!50] (u3) [above right of=u1] {$\tiny v$};
  \node[draw,circle,fill=gray!50] (u2) [below right of=u3] {$\tiny u$};
  \node[draw,fill=gray!50] (t) [right of=u2,node distance=1cm] {$\tiny t$};
  
  \path[->,thin,gray]
    (u1) edge (u2)
    (u1) edge[bend left] (u3)
    (u2) edge[bend left] (u3)
    (u2) edge (t)
    (u3) edge[bend left] (u2)
    (u3) edge[bend left] (u1)
    (u2) edge[bend left] (u1);
\path[->,line width=1.6pt,red] (u1) edge[bend left] (u3);
\end{tikzpicture}
\\ 
\hline
\hline
\begin{tikzpicture}[->,shorten >=1pt,auto,node distance=1.5cm,thin]
  \node[draw,circle,fill=gray!50] (u1) {$\tiny s$};
  \node[draw,circle,fill=blue!50] (u3) [above right of=u1] {$\tiny v$};
  \node[draw,circle,fill=blue!50] (u2) [below right of=u3] {$\tiny u$};
  \node[draw,fill=gray!50] (t) [right of=u2,node distance=1cm] {$\tiny t$};
  
  \path[->,thin,gray]
    (u1) edge (u2)
    (u1) edge[bend left] (u3)
    (u2) edge[bend left] (u3)
    (u2) edge (t)
    (u3) edge[bend left] (u2)
    (u3) edge[bend left] (u1)
    (u2) edge[bend left] (u1);
\path[->,line width=1.6pt,blue] (u2) edge[bend left] (u1)
								(u3) edge[bend left] (u1);
\end{tikzpicture}
&
\begin{tikzpicture}[->,shorten >=1pt,auto,node distance=1.5cm,thin]
  \node[draw,circle,fill=gray!50] (u1) {$\tiny s$};
  \node[draw,circle,fill=gray!50] (u3) [above right of=u1] {$\tiny v$};
  \node[draw,circle,fill=gray!50] (u2) [below right of=u3] {$\tiny u$};
  \node[draw,fill=gray!50] (t) [right of=u2,node distance=1cm] {$\tiny t$};
  
  \path[->,thin,gray]
    (u1) edge (u2)
    (u1) edge[bend left] (u3)
    (u2) edge[bend left] (u3)
    (u2) edge (t)
    (u3) edge[bend left] (u2)
    (u3) edge[bend left] (u1)
    (u2) edge[bend left] (u1);
\path[->,line width=1.6pt,red!50!blue] (u1) edge (u2)
								(u2) edge[bend left] (u1);
\node[above of=t,node distance=1cm] {$\footnotesize\begin{array}{c}\br \infty\\\bb \infty\end{array}$};
\end{tikzpicture}
&
\begin{tikzpicture}[->,shorten >=1pt,auto,node distance=1.5cm,thin]
  \node[draw,circle,fill=gray!50] (u1) {$\tiny s$};
  \node[draw,circle,fill=gray!50] (u3) [above right of=u1] {$\tiny v$};
  \node[draw,circle,fill=gray!50] (u2) [below right of=u3] {$\tiny u$};
  \node[draw,fill=gray!50] (t) [right of=u2,node distance=1cm] {$\tiny t$};
  
  \path[->,thin,gray]
    (u1) edge (u2)
    (u1) edge[bend left] (u3)
    (u2) edge[bend left] (u3)
    (u2) edge (t)
    (u3) edge[bend left] (u2)
    (u3) edge[bend left] (u1)
    (u2) edge[bend left] (u1);
\path[->,line width=1.6pt,red!50!blue] (u1) edge[bend left] (u3)
								(u3) edge[bend left] (u1);
\node[above of=t,node distance=1cm] {$\footnotesize\begin{array}{c}\underline{\br 0}\\\bb \infty\end{array}$};
\end{tikzpicture}
\\ 
\hline
\begin{tikzpicture}[->,shorten >=1pt,auto,node distance=1.5cm,thin]
  \node[draw,circle,fill=gray!50] (u1) {$\tiny s$};
  \node[draw,circle,fill=blue!50] (u3) [above right of=u1] {$\tiny v$};
  \node[draw,circle,fill=blue!50] (u2) [below right of=u3] {$\tiny u$};
  \node[draw,fill=gray!50] (t) [right of=u2,node distance=1cm] {$\tiny t$};
  
  \path[->,thin,gray]
    (u1) edge (u2)
    (u1) edge[bend left] (u3)
    (u2) edge[bend left] (u3)
    (u2) edge (t)
    (u3) edge[bend left] (u2)
    (u3) edge[bend left] (u1)
    (u2) edge[bend left] (u1);;
\path[->,line width=1.6pt,blue] (u2) edge[bend left] (u1)
								(u3) edge[bend left] (u2);

\end{tikzpicture}
&
\begin{tikzpicture}[->,shorten >=1pt,auto,node distance=1.5cm,thin]
  \node[draw,circle,fill=gray!50] (u1) {$\tiny s$};
  \node[draw,circle,fill=gray!50] (u3) [above right of=u1] {$\tiny v$};
  \node[draw,circle,fill=gray!50] (u2) [below right of=u3] {$\tiny u$};
  \node[draw,fill=gray!50] (t) [right of=u2,node distance=1cm] {$\tiny t$};
  
  \path[->,thin,gray]
    (u1) edge (u2)
    (u1) edge[bend left] (u3)
    (u2) edge[bend left] (u3)
    (u2) edge (t)
    (u3) edge[bend left] (u2)
    (u3) edge[bend left] (u1)
    (u2) edge[bend left] (u1);
\path[->,line width=1.6pt,red!50!blue] (u1) edge (u2)
								(u2) edge[bend left] (u1);
\node[above of=t,node distance=1cm] {$\footnotesize\begin{array}{c}\underline{\br \infty}\\\bb \infty\end{array}$};
\end{tikzpicture}
&
\begin{tikzpicture}[->,shorten >=1pt,auto,node distance=1.5cm,thin]
  \node[draw,circle,fill=gray!50] (u1) {$\tiny s$};
  \node[draw,circle,fill=gray!50] (u3) [above right of=u1] {$\tiny v$};
  \node[draw,circle,fill=gray!50] (u2) [below right of=u3] {$\tiny u$};
  \node[draw,fill=gray!50] (t) [right of=u2,node distance=1cm] {$\tiny t$};
  
  \path[->,thin,gray]
    (u1) edge (u2)
    (u1) edge[bend left] (u3)
    (u2) edge[bend left] (u3)
    (u2) edge (t)
    (u3) edge[bend left] (u2)
    (u3) edge[bend left] (u1)
    (u2) edge[bend left] (u1);
\path[->,line width=1.6pt,red!50!blue] (u1) edge[bend left] (u3)
								(u3) edge[bend left] (u2)
								(u2) edge[bend left] (u1);
\node[above of=t,node distance=1cm] {$\footnotesize\begin{array}{c}\underline{\br \infty}\\\bb \infty\end{array}$};
\end{tikzpicture}
\\ 
\hline
\begin{tikzpicture}[->,shorten >=1pt,auto,node distance=1.5cm,thin]
  \node[draw,circle,fill=gray!50] (u1) {$\tiny s$};
  \node[draw,circle,fill=blue!50] (u3) [above right of=u1] {$\tiny v$};
  \node[draw,circle,fill=blue!50] (u2) [below right of=u3] {$\tiny u$};
  \node[draw,fill=gray!50] (t) [right of=u2,node distance=1cm] {$\tiny t$};
  
  \path[->,thin,gray]
    (u1) edge (u2)
    (u1) edge[bend left] (u3)
    (u2) edge[bend left] (u3)
    (u2) edge (t)
    (u3) edge[bend left] (u2)
    (u3) edge[bend left] (u1)
    (u2) edge[bend left] (u1);;
\path[->,line width=1.6pt,blue] (u2) edge[bend left] (u3)
								(u3) edge[bend left] (u1);
\end{tikzpicture}
&
\begin{tikzpicture}[->,shorten >=1pt,auto,node distance=1.5cm,thin]
  \node[draw,circle,fill=gray!50] (u1) {$\tiny s$};
  \node[draw,circle,fill=gray!50] (u3) [above right of=u1] {$\tiny v$};
  \node[draw,circle,fill=gray!50] (u2) [below right of=u3] {$\tiny u$};
  \node[draw,fill=gray!50] (t) [right of=u2,node distance=1cm] {$\tiny t$};
  
  \path[->,thin,gray]
    (u1) edge (u2)
    (u1) edge[bend left] (u3)
    (u2) edge[bend left] (u3)
    (u2) edge (t)
    (u3) edge[bend left] (u2)
    (u3) edge[bend left] (u1)
    (u2) edge[bend left] (u1);
\path[->,line width=1.6pt,red!50!blue] (u1) edge (u2)
								(u2) edge[bend left] (u3)
								(u3) edge[bend left] (u1);
\node[above of=t,node distance=1cm] {$\footnotesize\begin{array}{c}\br \infty\\\underline{\bb 0}\end{array}$};
\end{tikzpicture}
&
\begin{tikzpicture}[->,shorten >=1pt,auto,node distance=1.5cm,thin]
  \node[draw,circle,fill=gray!50] (u1) {$\tiny s$};
  \node[draw,circle,fill=gray!50] (u3) [above right of=u1] {$\tiny v$};
  \node[draw,circle,fill=gray!50] (u2) [below right of=u3] {$\tiny u$};
  \node[draw,fill=gray!50] (t) [right of=u2,node distance=1cm] {$\tiny t$};
  
  \path[->,thin,gray]
    (u1) edge (u2)
    (u1) edge[bend left] (u3)
    (u2) edge[bend left] (u3)
    (u2) edge (t)
    (u3) edge[bend left] (u2)
    (u3) edge[bend left] (u1)
    (u2) edge[bend left] (u1);
\path[->,line width=1.6pt,red!50!blue] (u1) edge[bend left] (u3)
								(u3) edge[bend left] (u1);
\node[above of=t,node distance=1cm] {$\footnotesize\begin{array}{c}\underline{\br 0}\\\bb \infty\end{array}$};
\end{tikzpicture}
\\ 
\hline
\begin{tikzpicture}[->,shorten >=1pt,auto,node distance=1.5cm,thin]
  \node[draw,circle,fill=gray!50] (u1) {$\tiny s$};
  \node[draw,circle,fill=blue!50] (u3) [above right of=u1] {$\tiny v$};
  \node[draw,circle,fill=blue!50] (u2) [below right of=u3] {$\tiny u$};
  \node[draw,fill=gray!50] (t) [right of=u2,node distance=1cm] {$\tiny t$};
  
  \path[->,thin,gray]
    (u1) edge (u2)
    (u1) edge[bend left] (u3)
    (u2) edge[bend left] (u3)
    (u2) edge (t)
    (u3) edge[bend left] (u2)
    (u3) edge[bend left] (u1)
    (u2) edge[bend left] (u1);;
\path[->,line width=1.6pt,blue] (u2) edge[bend left] (u3)
								(u3) edge[bend left] (u2);
\end{tikzpicture}
&
\begin{tikzpicture}[->,shorten >=1pt,auto,node distance=1.5cm,thin]
  \node[draw,circle,fill=gray!50] (u1) {$\tiny s$};
  \node[draw,circle,fill=gray!50] (u3) [above right of=u1] {$\tiny v$};
  \node[draw,circle,fill=gray!50] (u2) [below right of=u3] {$\tiny u$};
  \node[draw,fill=gray!50] (t) [right of=u2,node distance=1cm] {$\tiny t$};
  
  \path[->,thin,gray]
    (u1) edge (u2)
    (u1) edge[bend left] (u3)
    (u2) edge[bend left] (u3)
    (u2) edge (t)
    (u3) edge[bend left] (u2)
    (u3) edge[bend left] (u1)
    (u2) edge[bend left] (u1);
\path[->,line width=1.6pt,red!50!blue] (u1) edge (u2)
								(u2) edge[bend left] (u3)
								(u3) edge[bend left] (u2);
\node[above of=t,node distance=1cm] {$\footnotesize\begin{array}{c}\underline{\br \infty}\\\bb \infty\end{array}$};
\end{tikzpicture}
&
\begin{tikzpicture}[->,shorten >=1pt,auto,node distance=1.5cm,thin]
  \node[draw,circle,fill=gray!50] (u1) {$\tiny s$};
  \node[draw,circle,fill=gray!50] (u3) [above right of=u1] {$\tiny v$};
  \node[draw,circle,fill=gray!50] (u2) [below right of=u3] {$\tiny u$};
  \node[draw,fill=gray!50] (t) [right of=u2,node distance=1cm] {$\tiny t$};
  
  \path[->,thin,gray]
    (u1) edge (u2)
    (u1) edge[bend left] (u3)
    (u2) edge[bend left] (u3)
    (u2) edge (t)
    (u3) edge[bend left] (u2)
    (u3) edge[bend left] (u1)
    (u2) edge[bend left] (u1);
\path[->,line width=1.6pt,red!50!blue] (u1) edge[bend left] (u3)
								(u2) edge[bend left] (u3)
								(u3) edge[bend left] (u2);
\node[above of=t,node distance=1cm] {$\footnotesize\begin{array}{c}\underline{\br \infty}\\\bb \infty\end{array}$};
\end{tikzpicture}
\\ 
\hline
\begin{tikzpicture}[->,shorten >=1pt,auto,node distance=1.5cm,thin]
  \node[draw,circle,fill=gray!50] (u1) {$\tiny s$};
  \node[draw,circle,fill=blue!50] (u3) [above right of=u1] {$\tiny v$};
  \node[draw,circle,fill=blue!50] (u2) [below right of=u3] {$\tiny u$};
  \node[draw,fill=gray!50] (t) [right of=u2,node distance=1cm] {$\tiny t$};
  
  \path[->,thin,gray]
    (u1) edge (u2)
    (u1) edge[bend left] (u3)
    (u2) edge[bend left] (u3)
    (u2) edge (t)
    (u3) edge[bend left] (u2)
    (u3) edge[bend left] (u1)
    (u2) edge[bend left] (u1);;
\path[->,line width=1.6pt,blue] (u2) edge (t)
								(u3) edge[bend left] (u1);
\end{tikzpicture}
&
\begin{tikzpicture}[->,shorten >=1pt,auto,node distance=1.5cm,thin]
  \node[draw,circle,fill=gray!50] (u1) {$\tiny s$};
  \node[draw,circle,fill=gray!50] (u3) [above right of=u1] {$\tiny v$};
  \node[draw,circle,fill=gray!50] (u2) [below right of=u3] {$\tiny u$};
  \node[draw,fill=gray!50] (t) [right of=u2,node distance=1cm] {$\tiny t$};
  
  \path[->,thin,gray]
    (u1) edge (u2)
    (u1) edge[bend left] (u3)
    (u2) edge[bend left] (u3)
    (u2) edge (t)
    (u3) edge[bend left] (u2)
    (u3) edge[bend left] (u1)
    (u2) edge[bend left] (u1);
\path[->,line width=1.6pt,red!50!blue] (u1) edge (u2)
								(u2) edge (t);
\node[above of=t,node distance=1cm] {$\footnotesize\begin{array}{c}\br 1\\\bb 1\end{array}$};
\end{tikzpicture}
&
\begin{tikzpicture}[->,shorten >=1pt,auto,node distance=1.5cm,thin]
  \node[draw,circle,fill=gray!50] (u1) {$\tiny s$};
  \node[draw,circle,fill=gray!50] (u3) [above right of=u1] {$\tiny v$};
  \node[draw,circle,fill=gray!50] (u2) [below right of=u3] {$\tiny u$};
  \node[draw,fill=gray!50] (t) [right of=u2,node distance=1cm] {$\tiny t$};
  
  \path[->,thin,gray]
    (u1) edge (u2)
    (u1) edge[bend left] (u3)
    (u2) edge[bend left] (u3)
    (u2) edge (t)
    (u3) edge[bend left] (u2)
    (u3) edge[bend left] (u1)
    (u2) edge[bend left] (u1);
\path[->,line width=1.6pt,red!50!blue] (u1) edge[bend left] (u3)
								(u3) edge[bend left] (u1);
\node[above of=t,node distance=1cm] {$\footnotesize\begin{array}{c}\underline{\br 0}\\\bb \infty\end{array}$};
\end{tikzpicture}
\\ 
\hline
\begin{tikzpicture}[->,shorten >=1pt,auto,node distance=1.5cm,thin]
  \node[draw,circle,fill=gray!50] (u1) {$\tiny s$};
  \node[draw,circle,fill=blue!50] (u3) [above right of=u1] {$\tiny v$};
  \node[draw,circle,fill=blue!50] (u2) [below right of=u3] {$\tiny u$};
  \node[draw,fill=gray!50] (t) [right of=u2,node distance=1cm] {$\tiny t$};
  
  \path[->,thin,gray]
    (u1) edge (u2)
    (u1) edge[bend left] (u3)
    (u2) edge[bend left] (u3)
    (u2) edge (t)
    (u3) edge[bend left] (u2)
    (u3) edge[bend left] (u1)
    (u2) edge[bend left] (u1);;
\path[->,line width=1.6pt,blue] (u2) edge (t)
								(u3) edge[bend left] (u2);
\end{tikzpicture}
&
\begin{tikzpicture}[->,shorten >=1pt,auto,node distance=1.5cm,thin]
  \node[draw,circle,fill=gray!50] (u1) {$\tiny s$};
  \node[draw,circle,fill=gray!50] (u3) [above right of=u1] {$\tiny v$};
  \node[draw,circle,fill=gray!50] (u2) [below right of=u3] {$\tiny u$};
  \node[draw,fill=gray!50] (t) [right of=u2,node distance=1cm] {$\tiny t$};
  
  \path[->,thin,gray]
    (u1) edge (u2)
    (u1) edge[bend left] (u3)
    (u2) edge[bend left] (u3)
    (u2) edge (t)
    (u3) edge[bend left] (u2)
    (u3) edge[bend left] (u1)
    (u2) edge[bend left] (u1);
\path[->,line width=1.6pt,red!50!blue] (u1) edge (u2)
								(u2) edge (t);
\node[above of=t,node distance=1cm] {$\footnotesize\begin{array}{c}\underline{\br 1}\\\bb 1\end{array}$};
\end{tikzpicture}
&
\begin{tikzpicture}[->,shorten >=1pt,auto,node distance=1.5cm,thin]
  \node[draw,circle,fill=gray!50] (u1) {$\tiny s$};
  \node[draw,circle,fill=gray!50] (u3) [above right of=u1] {$\tiny v$};
  \node[draw,circle,fill=gray!50] (u2) [below right of=u3] {$\tiny u$};
  \node[draw,fill=gray!50] (t) [right of=u2,node distance=1cm] {$\tiny t$};
  
  \path[->,thin,gray]
    (u1) edge (u2)
    (u1) edge[bend left] (u3)
    (u2) edge[bend left] (u3)
    (u2) edge (t)
    (u3) edge[bend left] (u2)
    (u3) edge[bend left] (u1)
    (u2) edge[bend left] (u1);
\path[->,line width=1.6pt,red!50!blue] (u1) edge[bend left] (u3)
								(u3) edge[bend left] (u2)
								(u2) edge (t);
\node[above of=t,node distance=1cm] {$\footnotesize\begin{array}{c}\br 2\\\underline{\bb 3}\end{array}$};
\end{tikzpicture}
\\ 
\hline
\hline
\end{tabular}
\caption{Normal form of the example given in Figure \ref{fig1-p}. The columns represent the two possible strategies of player \textsc{\color{red}Red} and the rows represent the six possible strategies of player \textsc{\color{blue}Blue}. Both players are minimizing their respective costs, indicated in the upper right corner in each cell. Since no situation is simultaneously row and column minimizing, this example has no NE. 
\label{fig4}}
\end{figure}

Finally we show an example for a $2$-person positive edge-symmetric shortest path game that has no UNE. This game, $\Gamma^{6s}$ is shown in Figure \ref{fig-G6s} is derived from the terminal game $\Gamma^6$ shown earlier. Note first that the counter clockwise moves have a large length for both players, larger than the length of any terminating move. This implies that in any UNE these moves will not be used. Let us also observe that from any position $v$ and for any two terminals $a_i$ and $a_j$, $i\neq j$ the $i(v)$-lengths of the paths from $v$ to these terminals (via the clockwise moves) compare exactly the same way as in the terminal game $\Gamma^6$: the $i(v)$-length of the $v\to a_i$ path is shorter than the $i(v)$-length of the $v\to a_j$ path if and only if player $i(v)$ prefers terminal $a_i$ to $a_j$ in $\Gamma^6$. Thus, any UNE in this shortest path game would correspond to a UNE in $\Gamma^6$, and \cite{BEGM12} proved that no such UNE exist. 

\section{Open Problems}

Two important questions concerning the existence of NE remain open. The first one is about terminal games.

\smallskip

\noindent{\bf (Q1)}
Does an $n$-person terminal game satisfying condition (CIW) have a NE?

\medskip

In this paper question {\bf (Q1)} is answered in the positive
for games on symmetric digraphs. Moreover,
in this case  condition  (CIW)  is not essential.

It was shown in \cite{BGMOV18} that in general
(for non-symmetric digraphs)
condition (CIW) is essential when $n > 2$,
while for $n=2$  (CIW)  is not needed \cite{BG03}.
Interestingly, if  the answer to  {\bf (Q1)} were negative
then paradoxically there should be a terminal game satisfying (CIW)  
in which all NE are infinite, that is, realized by infinite plays;
see \cite{BGMOV18}.

\bigskip

The second open problem is about shortest path games.

\smallskip

\noindent {\bf (Q2)}
Does a positive $2$-person shortest path game have a NE?

\medskip

The answer is negative for more than two players, see  \cite{GO14}.
In the present paper it is shown that
the answer is negative for general edge-symmetric $2$-person games, 
while it is positive for edge-symmetric positive games with any number of players.

\bigskip

{\bf Acknowledgements}
The third and forth authors were working within the framework of the
HSE University Basic Research Program. The fourth author was
supported in part by the state assignment topic no. 0063-2016-0003.

\end{document}